\documentclass[10pt, twocolumn, a4paper]{article}
\usepackage{titlesec} 

\usepackage[runin]{abstract} 

\abslabeldelim{---}                                     
\usepackage[top=2cm, bottom=2cm, left=1.5cm, right=1.5cm]{geometry}

\usepackage{amsmath}

\usepackage{newtxtext, newtxmath}

\usepackage{cite}
\usepackage{graphicx}
\usepackage{algorithmic}
\usepackage{textcomp}
\usepackage{xcolor}
\usepackage{tcolorbox} 

\usepackage{subcaption}

\usepackage{booktabs}   
\usepackage{multirow}   
\usepackage{listings}
\usepackage{siunitx}
\usepackage{acro}

\DeclareAcronym{1t1c}{
  short = 1T1C ,
  long  = One Transistor and One Capacitor
}

\DeclareAcronym{ACL}{
  short = ACL ,
  long  = Access Control List
}
\DeclareAcronym{ACS}{
  short = ACS ,
  long  = access control system
}
\DeclareAcronym{AD}{
  short = AD ,
  long  = associated data
}
\DeclareAcronym{AEAD}{
  short = AEAD ,
  long  = Authenticated Encryption with Associated Data
}
\DeclareAcronym{AES}{
  short = AES ,
  long  = Advanced Encryption Standard
}
\DeclareAcronym{AS}{
  short = AS ,
  long  = Authentication Server
}
\DeclareAcronym{AP}{
  short = AP ,
  long  = Access Point
}
\DeclareAcronym{alu}{
  short = ALU ,
  long  = Arithmetic Logic Unit
}

\DeclareAcronym{BER}{
  short = BER ,
  long  = bit error rate
}
\DeclareAcronym{bl}{
  short = BL ,
  long  = Bit Line
}

\DeclareAcronym{ca}{
  short = CA ,
  long  = Certification Authority
}
\DeclareAcronym{COTS}{
  short = COTS ,
  long  = Commercial Off-the-Shelf
}
\DeclareAcronym{CSI}{
  short = CSI ,
  long  = channel state information
}
\DeclareAcronym{cmos}{
  short = CMOS ,
  long  = Complementary Metal-Oxide-Semiconductor
}
\DeclareAcronym{u}{
  short = $\mu$ ,
  long  = carrier mobility
}
\DeclareAcronym{cl}{
  short = CL ,
  long  = Column Address Strobe Latency
}
\DeclareAcronym{crp}{
  short = CRP ,
  long  = Challenge-Response Pair
}
\DeclareAcronym{cpu}{
  short = CPU ,
  long  = Central Processing Unit
}
\DeclareAcronym{CDF}{
  short = CDF ,
  long  = Cumulative Distribution Function
}

\DeclareAcronym{DLP}{
  short = DLP ,
  long  = Discrete Logarithm Problem
}
\DeclareAcronym{DH}{
  short = DH ,
  long  = Diffie-Hellman
}
\DeclareAcronym{DoS}{
  short = DoS ,
  long  = Denial-of-Service
}
\DeclareAcronym{dram}{
  short = DRAM ,
  long  = Dynamic Random Access Memory
}
\DeclareAcronym{dut}{
  short = DUT ,
  long  = Device Under Test
}
\DeclareAcronym{ddr}{
  short = DDR ,
  long  = Double Data Rate
}
\DeclareAcronym{ded}{
  short = DED ,
  long  = Double Error Detection
}

\DeclareAcronym{eq}{
  short = EQ ,
  long  = Equilibration Circuit
}
\DeclareAcronym{ec}{
  short = EC ,
  long  = Error Correction
}
\DeclareAcronym{ecc}{
  short = ECC ,
  long  = Error-Correcting Code
}
\DeclareAcronym{EER}{
  short = EER ,
  long  = Equal Error Rate
}

\DeclareAcronym{fpga}{
  short = FPGA ,
  long  = Field Programmable Gate Array
}
\DeclareAcronym{FAR}{
  short = FAR ,
  long  = False Acceptance Rate
}
\DeclareAcronym{FRR}{
  short = FRR ,
  long  = False Rejection Rate
}

\DeclareAcronym{GCM}{
  short = GCM ,
  long  = Galois Counter Mode
}

\DeclareAcronym{HTTP}{
  short = HTTP ,
  long  = Hypertext Transfer Protocol
}
\DeclareAcronym{hd}{
  short = HD ,
  long  = Hamming Distance
}
\DeclareAcronym{hw}{
  short = HW ,
  long  = Hamming Weight
}
\DeclareAcronym{hdintra}{
  short = $\text{HD}_{\text{intra}}$ ,
  long  = intra-Hamming Distance
}
\DeclareAcronym{hdinter}{
  short = $\text{HD}_{\text{inter}}$ ,
  long  = inter-Hamming Distance
}
\DeclareAcronym{hc}{
  short = HC ,
  long  = Hamming code
}
\DeclareAcronym{HD}{
  short = HD ,
  long  = Hamming-Distance
}
\DeclareAcronym{HW}{
  short = HW ,
  long  = Hamming-Weight
}

\DeclareAcronym{IBE}{
  short = IBE ,
  long  = Identity Based Encryption
}
\DeclareAcronym{IC}{
  short = IC ,
  long  = integrated circuit
}
\DeclareAcronym{ICMP}{
  short = ICMP ,
  long  = Internet Control Message Protocol
}
\DeclareAcronym{ICS}{
  short = ICS ,
  long  = Industrial Control System
}
\DeclareAcronym{IDS}{
  short = IDS ,
  long  = Intrusion Detection System
}
\DeclareAcronym{IETF}{
  short = IETF ,
  long  = Internet Engineering Task Force
}
\DeclareAcronym{iiot}{
  short = IIoT ,
  long  = Industrial Internet of Things
}
\DeclareAcronym{IHS}{
  short = IHS ,
  long  = integrated heat spreader
}
\DeclareAcronym{IoT}{
  short = IoT ,
  long  = Internet of Things
}
\DeclareAcronym{IT}{
  short = IT ,
  long  = Information Technology
}
\DeclareAcronym{IV}{
  short = IV ,
  long  = initialization vector
}
\DeclareAcronym{ids}{
  short = $I_{\text{DS}}$ ,
  long  = Drain Source Current
}
\DeclareAcronym{idssub}{
  short = $I_{DSsub}$ ,
  long  = Sub-Threshold Drain Source Current
}
\DeclareAcronym{iqr}{
  short = IQR ,
  long  = interquartile range
}

\DeclareAcronym{KDF}{
  short = KDF ,
  long  = Key Derivation Function
}

\DeclareAcronym{LAN}{
  short = LAN ,
  long  = Local Area Network
}
\DeclareAcronym{LCD}{
  short = LCD ,
  long  = Liquid Crystal Display
}

\DeclareAcronym{MAC}{
  short = MAC ,
  long  = Medium Access Control
}
\DeclareAcronym{MANET}{
  short = MANET ,
  long  = Mobile Ad-hoc Network
}
\DeclareAcronym{MCU}{
  short = MCU ,
  long  = Microcontroller Unit
}
\DeclareAcronym{ML}{
  short = ML ,
  long  = Machine Learning
}
\DeclareAcronym{MRAM}{
  short = MRAM ,
  long  = Magnetoresistive Random Access Memory
}
\DeclareAcronym{mosfet}{
  short = MOSFET ,
  long  = Metal-oxide-Semiconductor Field-Effect Transistor
}
\DeclareAcronym{mv}{
  short = MV ,
  long  = Majority Voting
}

\DeclareAcronym{NFC}{
  short = NFC ,
  long  = Near Field Communication
}
\DeclareAcronym{NIC}{
  short = NIC ,
  long  = network interface card
}
\DeclareAcronym{nvs}{
  short = NVS ,
  long  = Non-Volatile Storage
}

\DeclareAcronym{OTP}{
  short = OTP ,
  long  = one-time-pad
}
\DeclareAcronym{OUI}{
  short = OUI ,
  long  = Organizational Unique Identifier
}

\DeclareAcronym{p2p}{
  short = P2P ,
  long  = Peer-to-Peer
}
\DeclareAcronym{PCP}{
  short = PCP ,
  long  = Priority Code Point
}
\DeclareAcronym{PKI}{
  short = PKI ,
  long  = Public-Key Infrastructure
}
\DeclareAcronym{PLC}{
  short = PLC ,
  long  = Power-Line Communication
}
\DeclareAcronym{PoC}{
  short = PoC ,
  long  = Proof-of-Concept
}
\DeclareAcronym{PRNG}{
  short = PRNG ,
  long  = Pseudo Random Number Generator
}
\DeclareAcronym{puf}{
  short = PUF ,
  long  = Physically Unclonable Function
}
\DeclareAcronym{pl}{
  short = PL ,
  long  = Programmable Logic
}
\DeclareAcronym{pcb}{
  short = PCB ,
  long  = Printed Circuit Board
}
\DeclareAcronym{pre}{
  short = PRE ,
  long  = Precharge Command
}
\DeclareAcronym{PPF}{
  short = PPF ,
  long  = Percent Point Function
}

\DeclareAcronym{QR}{
  short = QR ,
  long  = Quick Response
}
\DeclareAcronym{QoS}{
  short = QoS ,
  long  = Quality of Service
}

\DeclareAcronym{RAM}{
  short = RAM ,
  long  = Random Access Memory
}
\DeclareAcronym{RF}{
  short = RF ,
  long  = Radio Frequency
}
\DeclareAcronym{RNG}{
  short = RNG ,
  long  = Random Number Generator
}
\DeclareAcronym{RO}{
  short = RO ,
  long  = ring-oscillator
}
\DeclareAcronym{rtc}{
  short = RTC ,
  long  = Real-Time Clock
}
\DeclareAcronym{ROC}{
  short = ROC ,
  long  = Receiver Operating Characteristic
}

\DeclareAcronym{SDN}{
  short = SDN ,
  long  = Software Defined Network
}
\DeclareAcronym{SHA}{
  short = SHA ,
  long  = Secure Hash Algorithm
}
\DeclareAcronym{SME}{
  short = SME ,
  long  = Small and Medium Enterprise
}
\DeclareAcronym{SoC}{
  short = SoC ,
  long  = System on Chip
}
\DeclareAcronym{sram}{
  short = SRAM ,
  long  = Static Random Access Memory
}
\DeclareAcronym{spice}{
  short = SPICE ,
  long  = Simulation Program with Integrated Circuit Emphasis
}
\DeclareAcronym{senamp}{
  short = SenseAmp ,
  long  = Sense Amplifier
}
\DeclareAcronym{sodimm}{
  short = SO-DIMM ,
  long  = Small Outline Dual Inline Memory Module
}
\DeclareAcronym{sdram}{
  short = SDRAM ,
  long  = Synchronous Dynamic Random Access Memory
}
\DeclareAcronym{sec}{
  short = SEC ,
  long  = Single Error Correction
}
\DeclareAcronym{secded}{
  short = SECDED ,
  long  = {Single Error Correction, Double Error Detection}
}
\DeclareAcronym{smv}{
  short = SMV ,
  long  = Spatial Majority Voting
}

\DeclareAcronym{TRNG}{
  short = TRNG ,
  long  = true random number generator
}
\DeclareAcronym{vth}{
  short = $V_{\text{th}}$ ,
  long  = Threshold Voltage
}
\DeclareAcronym{twr}{
  short = tWR ,
  long  = Write Recovery Time
}
\DeclareAcronym{TCP}{
  short = TCP ,
  long  = Transmission Control Protocol
}
\DeclareAcronym{tmv}{
  short = TMV ,
  long  = Temporal Majority Voting
}

\DeclareAcronym{vdd}{
  short = $V_{\text{DD}}$ ,
  long  = Supply Voltage
}

\DeclareAcronym{wlan}{
  short = WLAN ,
  long  = Wireless Local Area Network
}
\DeclareAcronym{wl}{
  short = WL ,
  long  = Word Line
}
\usepackage[colorlinks=true, linkcolor=blue, citecolor=blue, urlcolor=blue]{hyperref}

\usepackage[utf8]{inputenc}
\usepackage[T1]{fontenc}

\let\OLDthebibliography\thebibliography
\renewcommand\thebibliography[1]{
  \OLDthebibliography{#1}
  \setlength{\parskip}{0pt}
  \setlength{\itemsep}{0pt plus 0.3ex}
}

\definecolor{paperBg}{RGB}{248, 248, 248}   
\definecolor{academicRed}{RGB}{215, 40, 60}  
\definecolor{academicTeal}{RGB}{0, 90, 220}       
\definecolor{academicGreen}{RGB}{15, 150, 50}  
\definecolor{academicGold}{RGB}{225, 95, 0}   

\lstdefinestyle{PaperC_contrast}{
    language=C,
    backgroundcolor=\color{paperBg},
    basicstyle=\ttfamily\footnotesize\color{black}, 
    keywordstyle=\color{academicRed}\itshape,
    keywordstyle=[2]\color{academicTeal}\itshape,
    morekeywords=[2]{uint8_t},
    keywordstyle=[3]\color{academicGreen},
    morekeywords=[3]{__attribute__, section},
    keywordstyle=[4]\color{academicTeal},
    morekeywords=[4]{CONFIG_PUF_SIZE},
    stringstyle=\color{academicGold},
    breaklines=true,
    showstringspaces=false,
    frame=single,
    rulecolor=\color{lightgray!60}, 
    xleftmargin=5pt, 
    framexleftmargin=3pt,
    aboveskip=\smallskipamount,
    belowskip=\smallskipamount,
}

\newenvironment{keywords}
    {\vspace{1ex}\noindent{\textbf{\textit{Index Terms---}}}}
    {\vspace{1ex}}

\usepackage{authblk}
\usepackage{microtype}
\usepackage{xurl}

\begin{document}

\title{\huge \bfseries Secure Authentication in Wireless IoT: Hamming Code Assisted SRAM PUF as Device Fingerprint}

\renewcommand*{\Authfont}{\fontsize{11}{13}\bfseries\selectfont} 

\renewcommand*{\Affilfont}{\normalsize\normalfont}

\author[1]{Florian Lehn}
\author[1]{Pascal Ahr}
\author[1, 2]{Hans D. Schotten}

\affil[1]{German Research Center for Artificial Intelligence, Germany}
\affil[2]{Division of Wireless Communications and Radio Positioning, RPTU University Kaiserslautern-Landau, Germany}

\affil[ ]{\textit{Email: \{florian.lehn, pascal.ahr, hans.schotten\}@dfki.de}}

\date{\vspace{1ex} \normalsize \textit{Please note: This is a preprint submitted to arXiv, licensed under arXiv.org perpetual, non-exclusive license. \\
This work is accepted but not yet published at the 30th ITG-Symposium, Mobile Communications - Technologies and Applications in Osnabrueck, Germany.}}

\twocolumn[
  \begin{@twocolumnfalse}

    \maketitle
    \begin{abstract} 
\ac{sram} \acp{puf} make use of intrinsic manufacturing variations in memory cells to derive device-unique responses.
Employing such hardware-rooted fingerprints for authentication, this work demonstrates a threshold-based authentication proof of concept for constrained \ac{iiot} devices.
The proposed scheme can reliably cap the the post-authentication \ac{BER} below 1\,\%.
Inherent \ac{sram} \ac{puf} unreliability is addressed by a resource-efficient combination of \ac{hc} \ac{ec} and \ac{tmv}.
Increasing \ac{hc} redundancy or \ac{tmv} count significantly reduces the \ac{BER}, albeit with diminishing returns and increasingly prohibitive computational overhead.
Furthermore, this work quantifies the threshold gap between strict reliability and security constraints.
This gap is reframed as a design budget which enables the resource-aware calibration of the acceptance threshold, \ac{puf} response length, and stabilization technique, without violating designed-for error limits.
Larger responses make reliability optimizations increasingly obsolete.
This comparative analysis establishes a comprehensive design space for \ac{puf} \ac{ec}, guiding future implementations in balancing \ac{ec} quality against resource constraints such as computational demand, power consumption, and implementation complexity.
\end{abstract}
    
    \vspace{3ex} 
  \end{@twocolumnfalse}
]

 \acresetall

\section{Introduction}
\label{sec:introduction}
\noindent The fourth industrial revolution relies on Big Data and \ac{iiot}. 
While Big Data collects, stores, and processes large volumes of highly diverse data, the \ac{iiot} infrastructure serves as the primary source generating this data.
The combination of both technologies enables novel insights into industrial environments.
The \ac{iiot} is composed of numerous small, internet-connected sensor nodes.
Typically, those devices are cost-sensitive and highly resource-constrained, characterized by low computational power and a reliance on low-capacity batteries or even battery-less operation~\cite{9233424, 8326905}.
Due to these constraints, security remains a critical challenge in \ac{iiot} devices.
To secure communications, especially across wireless networks, robust authentication is essential.
Usually handled via conventional methods such as asymmetric cryptography, these methods can be compromised or are too resource-intensive for \ac{iiot} devices~\cite{10.1145/3591464, 8326905}. 
Consequently, \acp{puf} have been demonstrated to be a suitable alternative~\cite{4261134, 10.1145/3591464}.

\ac{sram} \acp{puf} derive device-unique responses \( \text R\) to a startup challenge \( \text C\), forming a \ac{crp} like a digital fingerprint. 
This fingerprint is extracted from intrinsic hardware imperfections caused by uncontrollable manufacturing variations.
A drawback of intrinsic \ac{sram} \acp{puf} is the imperfect reliability of \( \text R\).
This is caused by unstable \ac{sram} cells or environmental influences such as temperature or voltage fluctuations~\cite{4674345}. 

With two ESP32-S3 microcontrollers communicating via a \ac{wlan} \ac{TCP} point-to-point connection, this work provides a proof of concept that the unreliability inherent to \acp{puf}, which manifests in the \ac{BER} of a \ac{puf}-based authentication process, can be addressed by resource-efficient strategies like basic \ac{hc}, \ac{mv}, and an authentication acceptance threshold \(\tau_{\text{BER}}\).
Here, different \ac{sec} and \ac{secded} schemes are compared, for which parity bits are stored as helper data.

\acl{hc} serves as an \ac{ec} scheme based on XOR logic operations. 
These bitwise manipulations are directly performed by the microcontroller's \ac{alu}, thus making them both fast and low-power.
Consequently, \acp{hc} requires significantly lower computational overhead compared to more complex \acp{ec} such as polar codes, making them highly suitable for low-power \acp{MCU}~\cite{10908150}.

The rest of this paper is structured as follows:
First, Section~\ref{sec:sota} provides an overview of related work. 
Thereafter, Section~\ref{sec:setup} details the design of the \ac{sram} \ac{puf}-based authentication testbed.
Section~\ref{sec:evaluation} experimentally assesses authentication performance for various \ac{ec} and \ac{mv} configurations and associated \(\tau_{\text{BER}}\) calibrations. 
Finally, Section~\ref{sec:conclusion} summarizes findings and highlights future research directions.

\section{Related Work}
\label{sec:sota}
\noindent Authentication, especially in \ac{iiot} domains relying on resource-constrained devices, is one of the main applications of \ac{sram} \acp{puf} and has been extensively discussed in the literature \cite{9233424, 8367637, 8326905}. 
Despite advantages such as retrofit capability, zero additional hardware overhead, and having low power and cost requirements while still featuring strong security properties, \ac{sram} \acp{puf} tend to produce unreliable responses. 
To address this drawback, various stabilization concepts have been proposed.

\acp{ecc}, such as polar codes and \acp{hc}, focus on correcting bit flips of the \ac{puf} to provide a stable response. This post-processing is divided into phases, including the creation of helper data \cite{8254007, 10908150}.
Additionally, there are pre-selection and masking schemes to exclusively include stable bits in the response while discarding unstable ones \cite{11141806}, or in combination with an \ac{ecc}, such as \ac{hc}, to correct unstable bits \cite{10908150}.

Another approach involves stabilizing the response using \ac{mv}.
Instead of using the raw base response, there is a post-processing step that combines a set of responses. 
Differing in the voting dimension, there are two schemes: \ac{smv} and \ac{tmv}.
\ac{smv}, as in \cite{978-3-319-03491-1_3}, groups adjacent bit responses and selects the binary value inside the voting window with the highest occurrence or based on a certain threshold \cite{10733719}. In contrast, \ac{tmv} selects the binary value based on repeated responses for a specific bit location within the time dimension \cite{6224314}.

However, the landscape of these \ac{puf} stabilization techniques is fragmented, and a standardized combined approach is missing.
Additionally, while these stabilization and \ac{ec} techniques improve reliability, they can introduce prohibitive computational complexity, incur high energy overheads, require substantial \ac{nvs} to store helper data, or lead to significant entropy loss~\cite{10908150, 978-3-319-03491-1_3, 10733719}.
This introduces competing design trade-offs, of which there is a lack of comparative real-world analysis. 
Furthermore, applying strict authentication thresholds inherently forces a compromise between security and usability, creating a direct inverse relationship between the \ac{FAR} and \ac{FRR}~\cite{3705677.3705691}, 
which necessitates an application-specific calibration to balance these competing metrics, as well as critical resource limitations in \ac{iiot} devices.

Motivated by these constraints,
the main contribution of this work lies in combining a complexity-aware \ac{hc} with \ac{tmv} in a threshold-based authentication scheme. 
The objective is to mitigate the inherent instability of \ac{sram} \acp{puf} for authentication in \ac{iiot} while focusing on resource-efficient solutions.
Specifically, this requires assessing and balancing the associated computational and memory overheads, and calibrating the authentication threshold in a strictly resource and security-aware manner.
Consequently,
this work provides a comprehensive and comparative design space exploration for low-overhead \ac{puf}-assisted authentication primitives, 
yielding insights highly relevant for the application of resource-constrained \ac{iiot} devices in real-world environments.

\section{Setup for Evaluation}
\label{sec:setup}
\noindent The employed ESP32-S3 \acp{MCU} run FreeRTOS to enable wireless communication via the included Wi-Fi stack. 
To capture the \ac{sram} startup pattern reliably, the evaluated addresses must reside in a memory region that is never overwritten by the C runtime startup sequence. 
In the ESP-IDF toolchain, the runtime copies initialised variables from flash and zeroes the \texttt{.bss} segment before \texttt{app\_main} is called.
Therefore a \ac{puf} array is placed in the \texttt{.noinit} section using
the GCC attribute:

\begin{lstlisting}[style=PaperC_contrast]
static uint8_t puf_array[CONFIG_PUF_SIZE]
    __attribute__((section(".noinit")));
\end{lstlisting}

The linker assigns a fixed address to this region, ensuring that neither the startup code nor the \ac{nvs} flash subsystem ever accesses it. 
This approach is adopted, as it provides the strongest determinism guarantee. . 

\begin{figure}[!b]
    \centering
    \includegraphics[width=0.75\columnwidth]{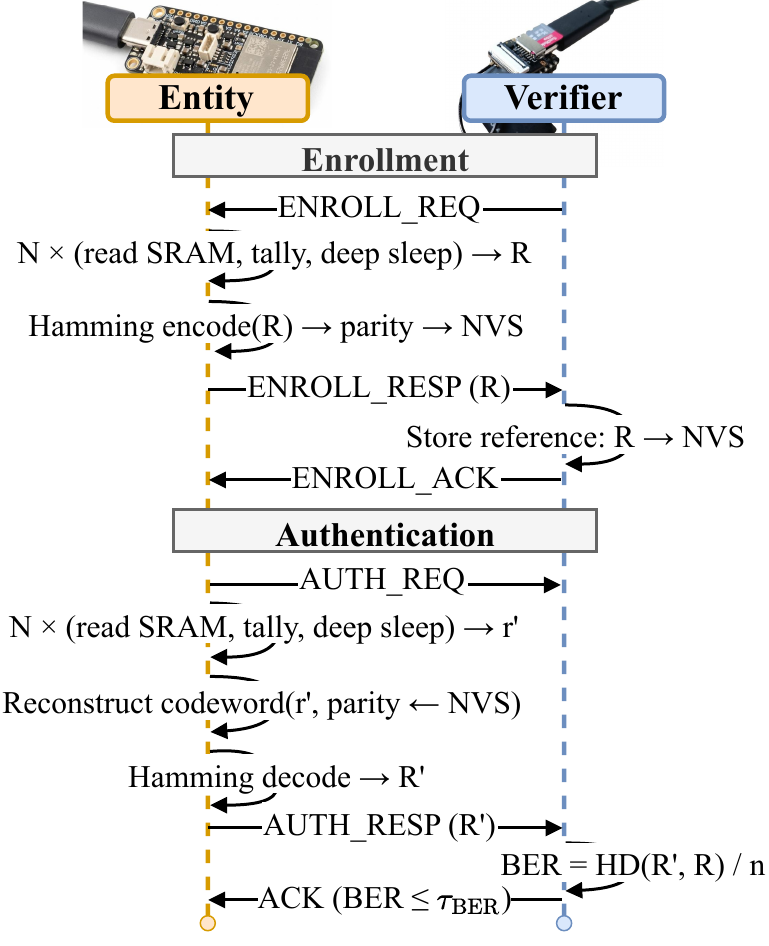}
    \caption{Sequence diagram of the \ac{puf}-based authentication protocol, illustrating the enrollment and authentication phases.}
    \label{fig:system_flow}
\end{figure}

The ESP32 family supports deep sleep. This power saving mode disconnects the main \ac{sram} and the \ac{cpu} from power, only the \ac{rtc} \ac{sram} remains powered for the hardware timers and wakeup logic.
\ac{rtc} memory makes it possible to store persistent data, like the \ac{mv} results. 
To implement \ac{tmv}, the ESP32 deep-sleeps between \ac{puf} readings to power-cycle main \ac{sram} and obtain statistically independent samples. 
The power-off time can be tailored to the \ac{sram} properties of the hardware in use via the ESP-IDF libraries.
The \ac{mv} occurs during system startup at the beginning of the main task, prior to Wi-Fi and other initializations. 
This sequence is necessary because deep sleep powers off the main \ac{cpu} and it's main \ac{sram}, thereby deleting the current execution state and resetting the entire system. 
The majority voted \ac{puf} response is then kept in memory for future authentication requests.
Crucially, the \ac{sram} memory is explicitly zeroed before each evaluation cycle before deep sleep to mitigate arbitrary \ac{sram} data remanence effects~\cite{8009192, 7366576}, preventing dependencies on previous execution states and prior \ac{puf} evaluations during the \ac{mv} process. 
This routine guarantees that the pre-power-down state of the memory is always identical, yielding highly reproducible \ac{puf} responses.
For \ac{ec}, 
helper data is generated during enrollment after \ac{mv} and stored in \ac{nvs} flash of the entity to persist across subsequent \ac{sram} power or deep sleep cycles. 
Alternatively, if the threat model requires it, \ac{ec} could similarly be performed by the verifier.
Fig.~\ref{fig:system_flow} highlights the \ac{sram} \ac{puf}-assisted authentication protocol as implemented in software on two ESP32-S3 \acp{MCU}. 

To clarify the computation overhead of \ac{mv} and \ac{ec}, the
techniques are implemented as follows: \ac{mv} acquires $N$ independent \ac{puf}
readings by power-cycling the ESP32 via configurable deep-sleep intervals. 
On each boot the raw \ac{sram} \ac{puf} response is immediately transformed into a per-bit counter stored
in \ac{rtc} memory, which persists across deep sleep.
Each counter accumulates $N$ passes of \ac{puf} size $n$ counter increments before thresholding
each counter against $N/2$ to produce the final bit string on the last boot.
For \ac{ec}, only the parity bits are persisted in \ac{nvs} as a
single packed byte per codeword, loaded as one contiguous byte array at authentication
time. The full Hamming codeword is then reconstructed on-the-fly by combining
the current raw \ac{puf} reading with the stored parity before syndrome-based
\ac{sec} or \ac{secded} is applied in a single pass over all codewords.

\section{Results of Evaluation}
\label{sec:evaluation}
\begin{figure}[htbp]
  \centering
  \includegraphics[width=\columnwidth]{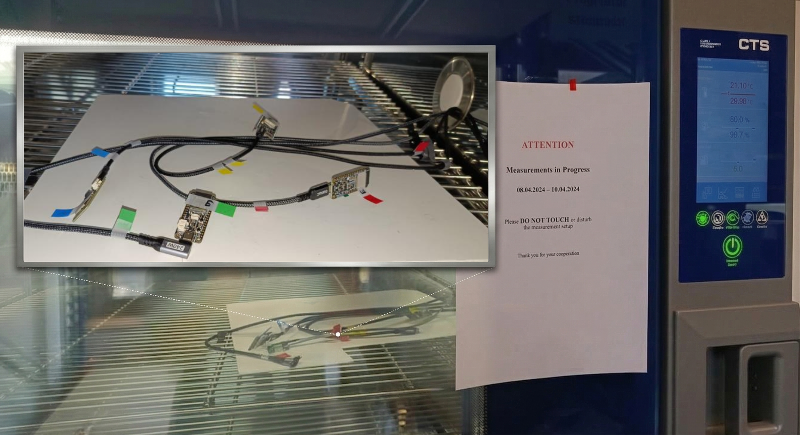}
  
  \caption{
  Experimental setup for measurements within a CTS environmental testing chamber. The composite view includes a smaller detailed inset photograph, showing four specific ESP32-S3 circuit boards with labeled connection cables during simultaneous testing. 
  }
  \label{fig:measurement_setup}
\end{figure}

\noindent Data collection was performed on a setup involving, in total, six ESP32-S3 entity \acp{MCU} running at 160\,MHz under controlled conditions in a climate chamber at \SI{21}{\celsius} and \SI{50}{\percent} relative humidity, configured as highlighted in Fig.~\ref{fig:system_flow}, with one verifier connected to one entity via a \ac{TCP} \ac{wlan} connection. 
The climate chamber setup is shown in Fig.~\ref{fig:measurement_setup}.
Based on preliminary testing the deep sleep time between \ac{puf} readings is configured to \SI{50}{\milli\second}.
On the verifier, measurements are logged to an onboard SD card. 
Each \ac{ecc} and \ac{mv} configuration gets its own separate enrollment and subsequent authentication sweep.
Due to evaluation time restrictions, each sweep is performed for $45$ iterations per device, with results aggregated across all entity devices into the same distribution plots.
All \ac{puf} responses are initially extracted at a full length of $2048$\,bits. To evaluate smaller \ac{puf} sizes $n$, these baseline responses are partitioned into $k = 2048/n_\text{collected}$ adjacent blocks, proportionally expanding the available sample pool for shorter response lengths. 
While this device and per-config iteration count is sufficient to validate the \ac{ecc} and \ac{mv} design space (see Section~\ref{subsec:Design_Space_Exploration}) and calibrate an error-constrained and resource-aware acceptance threshold (see Section~\ref{subsec:Threshold_Calibration}), the \ac{FAR} analysis will rely on an analytical impostor model rather than empirical inter-device measurements. 
Multi-board empirical validation of \acp{FAR} and other inter-device security metrics with a larger and diversified chip population remains future work.

\subsection{Design Space Exploration for Robust PUF Authentication}
\label{subsec:Design_Space_Exploration}
\noindent The subsequent design space exploration is essentially balancing three competing pillars: reliability, resource overhead (memory/\ac{cpu}/energy), and bit uniformity. 
This works reliability optimization strategy hinges on two core principles:
\begin{itemize}
    \item Helper Data Generation: Here the work discusses the overhead of storing parity bits for \ac{ec}, in this case \ac{hc}.
    \item Temporal Reliability: Using \ac{mv} (reading the \ac{sram} $N$ times) to filter out unstable bits. This work assesses a symmetric voting configuration, where the same amount of votes is applied during enrollment and authentication ($N_\text{enroll}=N_\text{auth}=N$).
\end{itemize}
For both cases the work also discusses the associated execution time overhead.

\subsubsection{Bit Uniformity}
\begin{figure}[!tb]
    \centering
    \begin{subfigure}[b]{0.49\linewidth} 
        \centering
        \includegraphics[width=\linewidth]{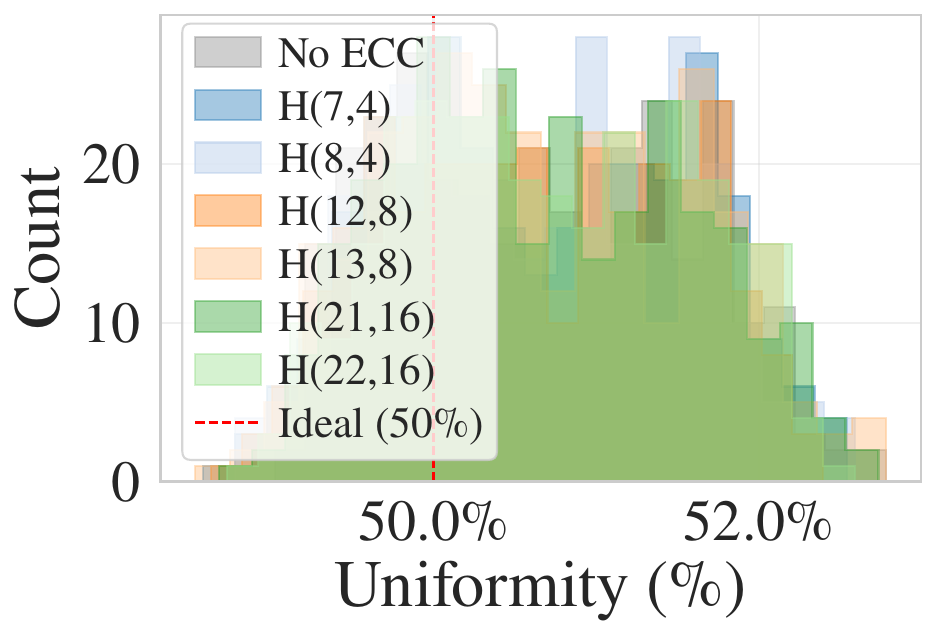} 
        \caption{
        Without \ac{mv} and \ac{ec}.
        } 
        \label{subfig:unif_hist_no_ecc}
    \end{subfigure}
    \hfill
    \begin{subfigure}[b]{0.49\linewidth} 
        \centering
        \includegraphics[width=\linewidth]{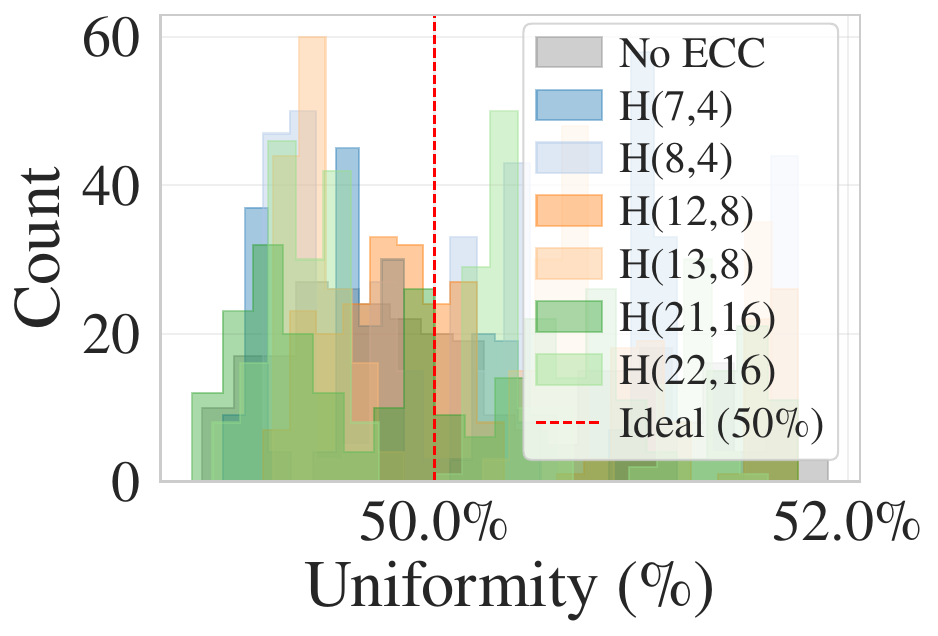}
        \caption{
        With \ac{mv}~($N=20$) and \ac{ec}.
        }
        \label{subfig:unif_hist_with_ecc}
    \end{subfigure}
    
    \caption{Uniformity distributions per measured \ac{ecc} scheme before (\subref{subfig:unif_hist_no_ecc}) and after (\subref{subfig:unif_hist_with_ecc}) \ac{puf} stabilization techniques.
    }
    \label{fig:unif_hist}
\end{figure}

Uniformity, following~\cite{Maiti2013, halak_physically_2018}, measures the bit bias in a \ac{puf} response as the fractional \ac{HW}.
In an ideal case this yields a uniform distribution of $\mathrm{Unif}\approx 50\%$. 
Fig.~\ref{fig:unif_hist} highlights that the evaluated \ac{sram} \ac{puf}-assisted authentication approach approximates this ideal. It also demonstrates that \ac{mv} and \ac{hc} \ac{ec} tighten the distribution, removing some of the inherent noise.

\subsubsection{Reliability}
\begin{figure}[!tb]
    \centering
    
    \begin{subfigure}{\columnwidth}
        \centering

        \includegraphics[width=\linewidth]{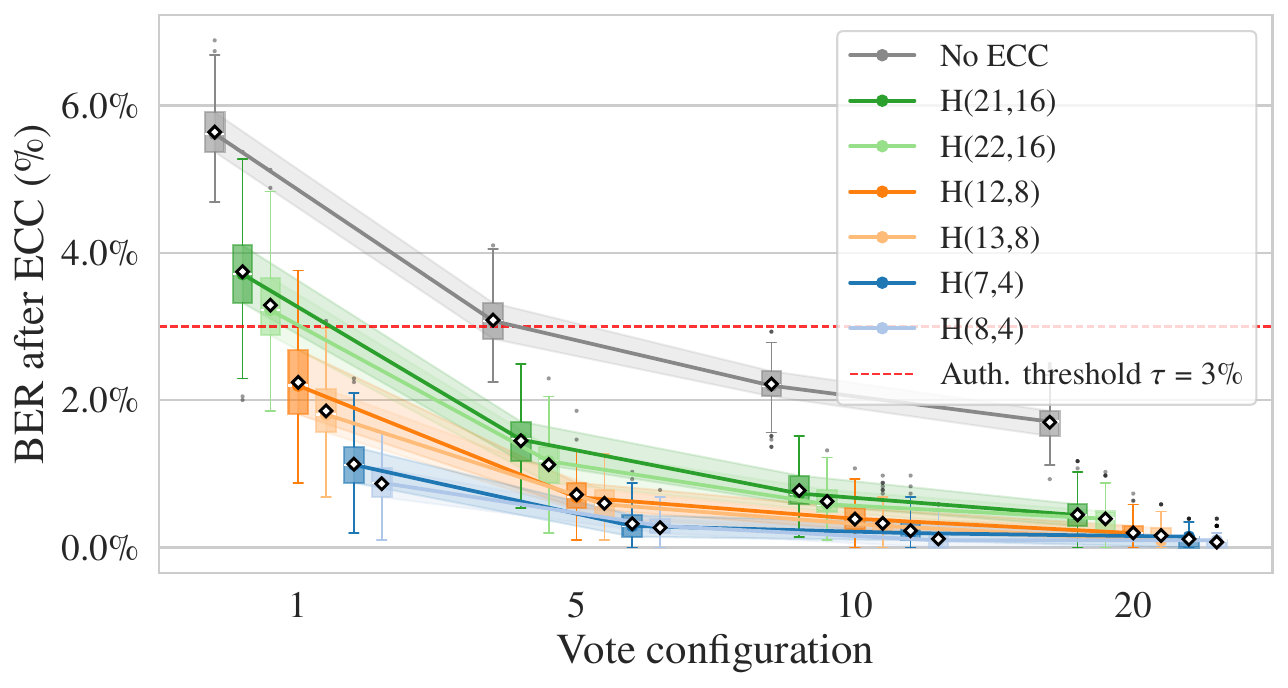}
        \caption{\ac{puf} response size $n=2048$.}
        \label{fig:ber_vs_votes_2048}
    \end{subfigure}
    
    \vspace{1em}
    
    \begin{subfigure}{\columnwidth}
        \centering
        \includegraphics[width=\linewidth]{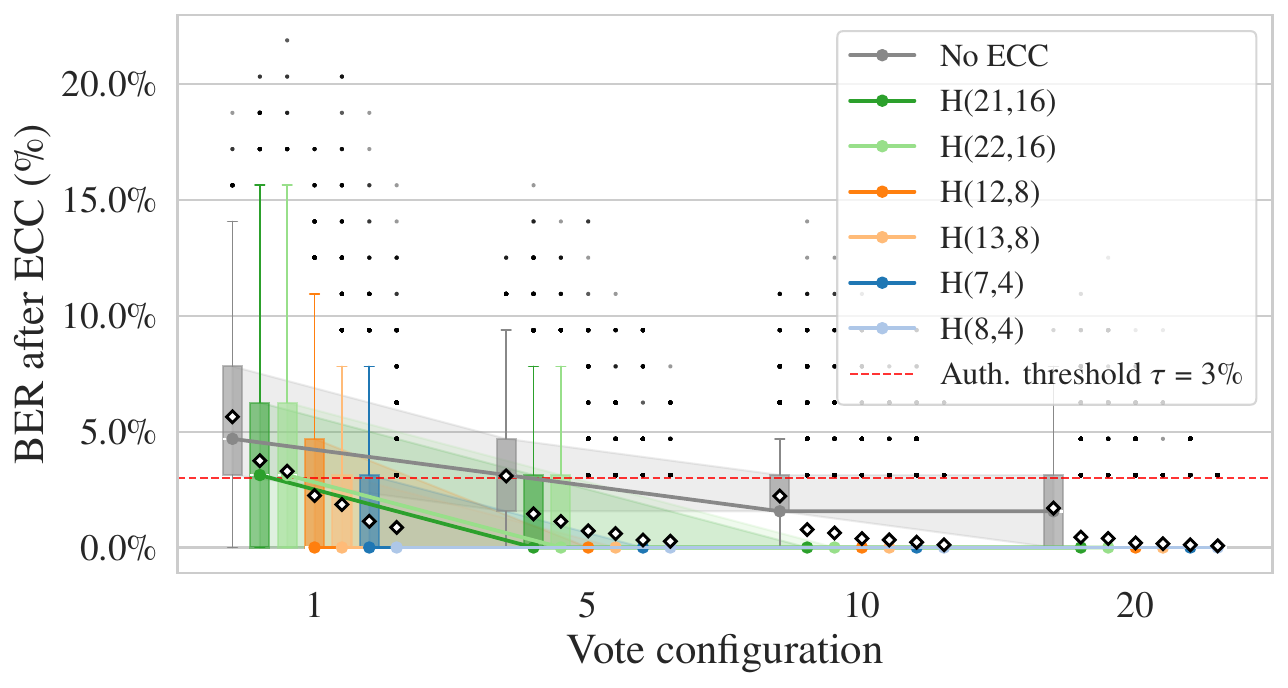}
        \caption{\ac{puf} response size $n=64$.}
        \label{fig:ber_vs_votes_64}
    \end{subfigure}
    
    \caption{Post-correction \ac{BER} vs. \ac{mv} count for each \ac{ec} scheme for~(\subref{fig:ber_vs_votes_64}) $n=64$ and~(\subref{fig:ber_vs_votes_2048}) $n=2048$ bits.
    Median trends are connected by lines, with shaded bands indicating the \ac{iqr}. Box-and-whisker distributions show the full per-iteration spread at each vote count, with whiskers extending to $1.5\times$\ac{iqr}. The dashed red line marks a potential \SI{3}{\percent} \(\tau_{\text{BER}}\) authentication threshold.}
    \label{fig:ber_vs_votes}
\end{figure}

The authentication \ac{BER} for different \ac{ec} configurations is presented in Fig.~\ref{fig:ber_vs_votes}
, which indicates the post-correction \ac{BER} as a function of \ac{mv} count $N$ for all evaluated \acp{ecc}.
The \ac{BER} calculation is based on the normalized \ac{HD}~\cite{halak_physically_2018} between the enrolled response and the response transmitted for authentication. 

All \ac{ec} variants exhibit a monotonic \ac{BER} reduction with increasing~$N$, but with diminishing returns, especially beyond $N \approx 5$. 
\ac{secded} provides a marginal mean improvement over standard \ac{sec} \acp{hc}. In a standard \ac{sec}-only code, a 2-bit error produces a nonzero syndrome (the calculated error pattern) that is indistinguishable from that of a single-bit error at a different position, which can be seen as a form of syndrome aliasing. 
This causes the decoder to miscorrect, flipping an additional bit and turning two errors into three. 
The extended parity bit in \ac{secded} detects this case, suppressing the miscorrection and leaving the two original errors uncorrected rather than creating a third.
By preventing such 3-bit miscorrections, \ac{secded} eliminates some of the higher \ac{BER} outliers seen in standard \ac{sec} codes, resulting in narrower \acp{iqr} in Fig.~\ref{fig:ber_vs_votes}.
Consequently, the mitigated syndrome aliasing phenomenon can be observed in the reduced dispersion of the \ac{secded} box plots.
In addition, for smaller \ac{puf} response sizes $n$, fewer codewords per response are formed, so a single miscorrected block has a disproportionate impact on the overall \ac{BER}, amplifying the outlier effect due to miscorrections, which is visible in Fig.~\ref{fig:ber_vs_votes} when comparing (\subref{fig:ber_vs_votes_64}) $n=64$ and~(\subref{fig:ber_vs_votes_2048}) $n=2048$ bits.
Therefore, as demonstrated in Fig.~\ref{fig:ber_vs_votes_64}, for small~$n$, the whiskers and outliers of the \ac{hc}-corrected \ac{BER} distributions, especially those that are \ac{sec}-only with low redundancy such as H(21,16) or H(12,8), can exceed those of the uncorrected baseline, confirming that miscorrection can degrade worst-case reliability beyond the no-\ac{ecc} case when only few codewords average out the effect. 
Nevertheless, the mean \ac{BER} across all \ac{mv} and \ac{ec} configurations (the diamond markers in Fig.~\ref{fig:ber_vs_votes}), stays the same when comparing lower $n$~(\ref{fig:ber_vs_votes_64}) and higher $n$~(\ref{fig:ber_vs_votes_2048}), only the statistical spread is altered.

Additionally, as expected, the \ac{BER} improves as the number of \ac{hc} data bits, so the code rate, decreases, 
with \(\mathrm{H}(7,4)\) configurations outperforming \(\mathrm{H}(12,8)\) 
and \(\mathrm{H}(21,16)\) variants. 
Taking into account fewer data bits per \ac{hc} correction block reduces the potential for $>1$~bit (\ac{sec}) or $>2$~bit (\ac{secded}) uncorrectable and undetectable bit flips and thereby false miscorrection attempts within a single codeword.

\subsubsection{Reliability vs. Memory Overhead}
\begin{figure}[tb]
    \centering
    \includegraphics[width=\columnwidth]{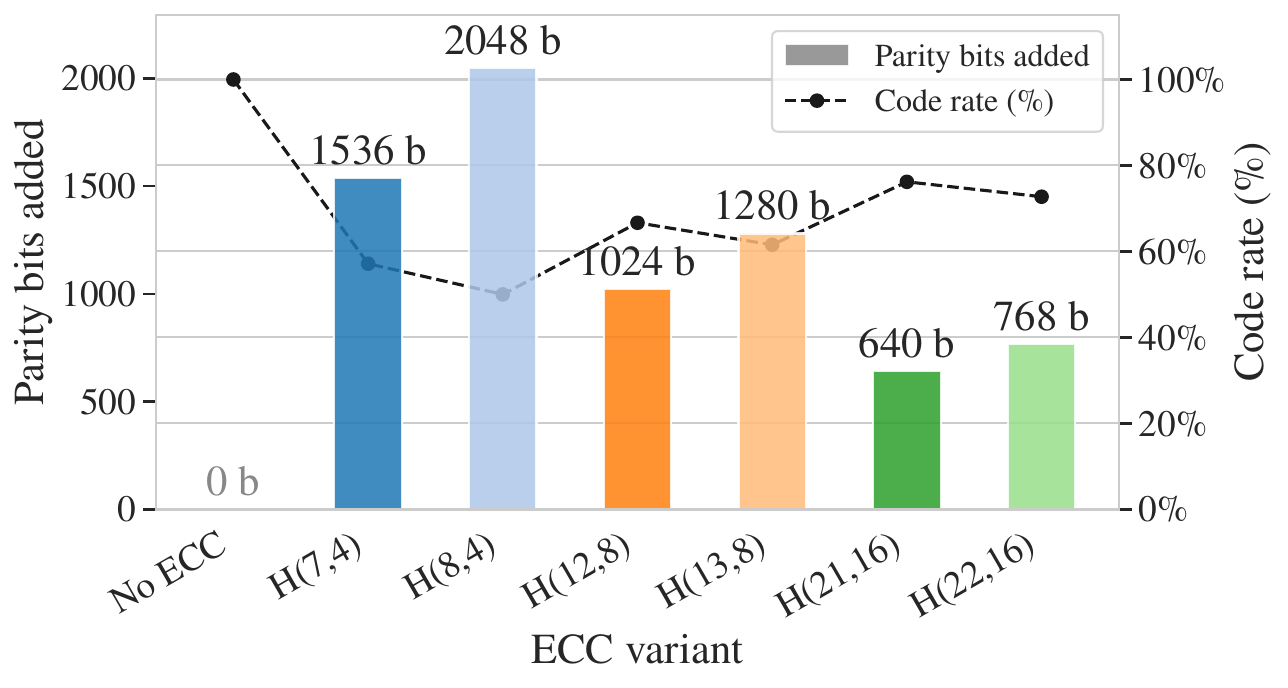}
    \caption{
Trade-off between \ac{nvs} overhead and data efficiency for each \ac{hc} variant for a 256-byte \ac{puf} response, where the code rate $R = n/k$ relates data bits~$n$ to codeword bits~$k$.
}

    \label{fig:parity_footprint}
\end{figure}

Designers must evaluate whether the suppression of error aliasing and the resulting reliability gains justify the increased helper data storage 
overhead of the additional \ac{secded} parity bit. 
In addition, the performance gain of fewer data bits per codeword comes at the cost of increased helper data overhead. 
Fig.~\ref{fig:parity_footprint} illustrates this \ac{ec} memory overhead,
with a higher code rate implying higher data efficiency but also worse reliability (see Fig.~\ref{fig:ber_vs_votes}).
Smaller correction blocks like H(7,4) impose a significantly lower data-to-parity ratio, forcing a trade-off between maximizing \ac{hc} \ac{ec} quality and minimizing the parity overhead. This reliability vs. memory overhead trade-off is is especially relevant for \ac{hc} implementations on resource-constrained devices like the ESP32 family with limited \ac{nvs} flash for helper data storage. 
Furthermore, the public transmission and storage of helper data introduce severe security vulnerabilities, as they can inadvertently leak information about the underlying secret key~\cite{8254007} or be exploited by machine learning models to predict \ac{puf} responses~\cite{10.1145/3591464}, imposing strict constraints on the assumed threat model.
Future real-world deployments must mitigate these advanced attack vectors, but such techniques remain beyond the scope of this work and readers are referred to related work~\cite{8254007, 10.1145/3591464}.

\subsubsection{Computational Demand}
The computational demand and execution time overhead is crucial in selecting the appropriate \ac{puf} stabilization technique. 
This demand can be characterized via the execution time overhead and is generally directly proportional to the energy demand of the operation.
This discounts some differences in the power demand of different architectural operations.
Data locality, for example, is important. Moving data by accessing memory often has a higher power demand than local arithmetic operations.
Increasing the \ac{mv} count has a high impact as it does not just increase \ac{cpu} time, but also increases the number of energy-demanding \ac{sram} read cycles. In this \ac{tmv} implementation specifically, it also creates repeated boot cycles per count. These repeated reads and especially boot cycles are expected to be the primary energy driver. 
Similarly, choosing a \ac{hc} with a higher memory overhead leads to increased energy-intensive \ac{nvs} flash transitions to retrieve stored helper data.
Nevertheless, without conducting detailed energy demand measurements, this work takes on the more focused approach of quantifying the general computational overhead via execution time measurements.

\begin{figure}[tb]
    \centering
    \begin{subfigure}[b]{\linewidth}
        \centering
        \includegraphics[width=\linewidth]{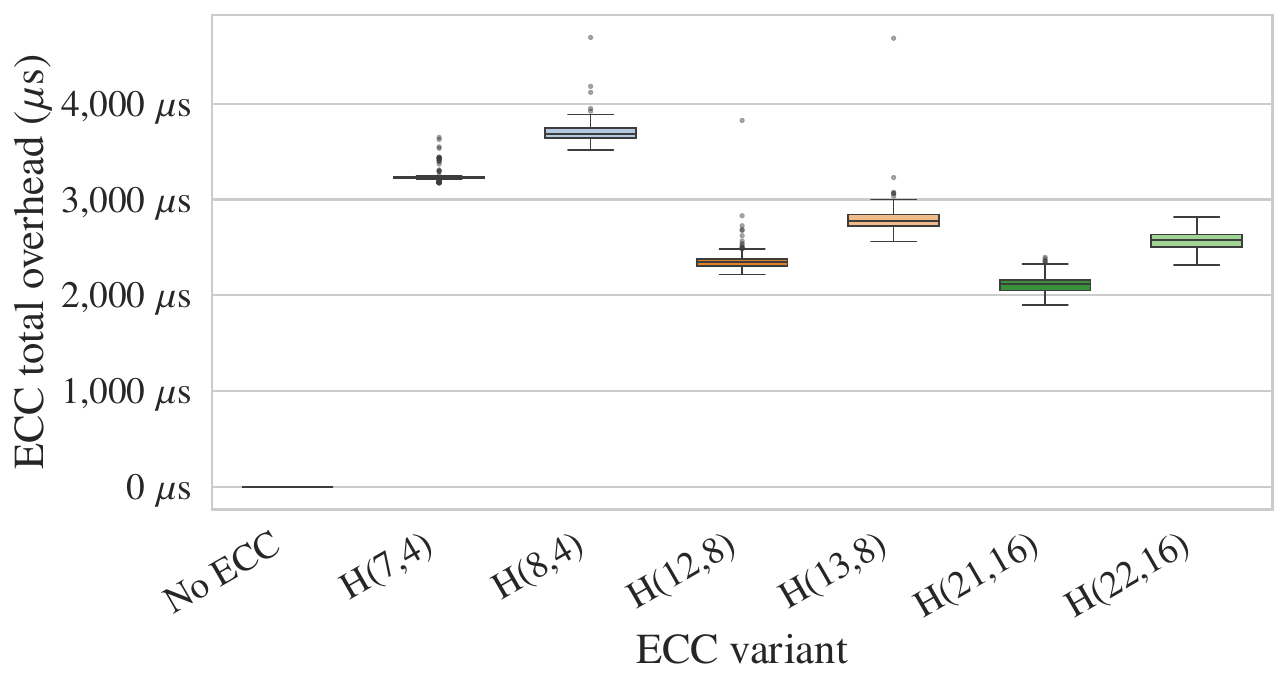}
        \caption{Boxplot distribution of \ac{hc} \ac{ec} execution time for each code variant without \ac{mv}. Variance reflects the number of bit errors requiring correction, which differs between iterations due to \ac{sram} noise. 
        \ac{ec} execution is composed of \ac{ec} computation time (syndrome computation and error correction) and memory overhead, mainly \ac{nvs} transactions to retrieve helper data.}
        \label{subfig:timing_breakdown_box}
    \end{subfigure}

    \vspace{0.5cm}

    \begin{subfigure}[b]{\linewidth}
        \centering
        \includegraphics[width=\linewidth]{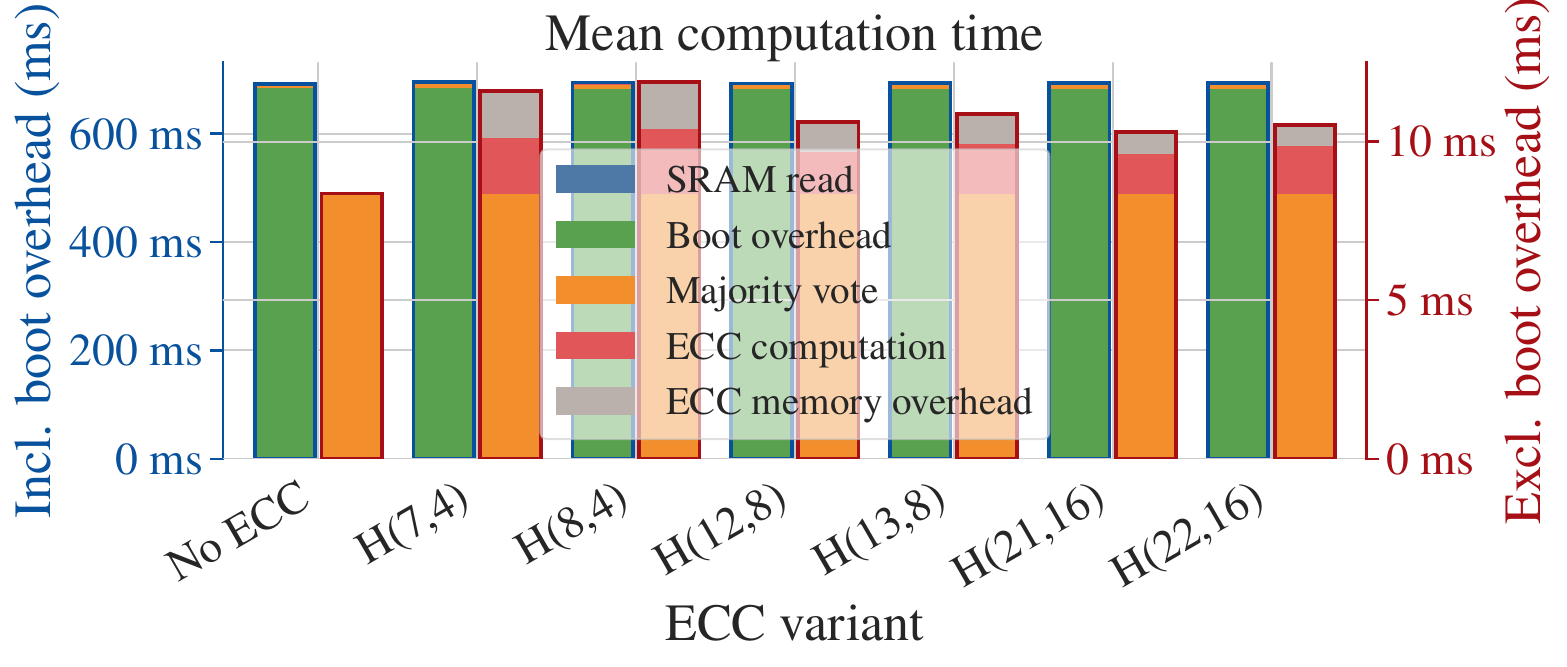}
        \caption{Mean on-device computation time per \ac{ec} variant for a \ac{mv} configuration of $N=10$. Times are decomposed into \ac{sram} readout (blue, negligible), \ac{mv} computation (orange), and \ac{ec} computation (red) and memory access overhead (grey). From the boot overhead (green) deep-sleep discharge intervals are excluded.}
        \label{subfig:timing_breakdown_bar}
    \end{subfigure}

    \caption{\ac{puf} authentication timing breakdown. (\subref{subfig:timing_breakdown_box}) shows the \ac{hc} \ac{ec} time distribution per variant. (\subref{subfig:timing_breakdown_bar}) shows that \ac{mv} dominates computation time, with \ac{ec} correction contributing a comparatively small, slightly variant-dependent overhead.}
    \label{fig:timing_breakdown}
\end{figure}

Fig.~\ref{fig:timing_breakdown} highlights the execution time trade-offs between different \ac{ec} codes and \ac{mv} vote counts.
While decreasing the code rate has little effect on pure computation time, the resulting parity overhead significantly increases \ac{nvs} read cycles and associated execution times, making these factors, alongside security implications~\cite{10.1145/3591464, 8254007} and correction quality, the primary drivers in selecting the appropriate \ac{hc} variant.
With increasing $N$, \ac{mv} exhibits a significant \ac{BER} improvement (see Fig.~\ref{fig:ber_vs_votes}) but also the most prohibitive impact on the execution time (see Fig.~\ref{subfig:timing_breakdown_bar}), especially due to the boot overhead which does not even capture the full reboot cost, because only the ESP-IDF software timer is used for measurement, and it is initialised only after the first-stage bootloader and the second-stage firmware loader from flash have already completed. 
Additionally, the power-off sleep times are excluded, which carry considerable latency but negligible energy overhead due to the low deep-sleep current draw.
Therefore, considering the diminishing returns in increasing $N$ (see Fig.~\ref{fig:ber_vs_votes}) but increasingly prohibitive overhead (see Fig.~\ref{subfig:timing_breakdown_bar}), 
selecting the right combination of \ac{mv} count $N$ and \ac{hc} code rate is paramount in meeting resource constraints and \ac{puf} reliability requirements simultaneously. 
Consequently, for a given reliability requirement, the \ac{mv} count $N$ and \ac{hc} code rate should be selected as small as possible.

\subsection{Resource-Aware Threshold Calibration for Authentication}
\label{subsec:Threshold_Calibration}
\begin{figure}[tb]
    \centering
    \begin{subfigure}[b]{\linewidth}
        \centering
        \includegraphics[width=\linewidth]{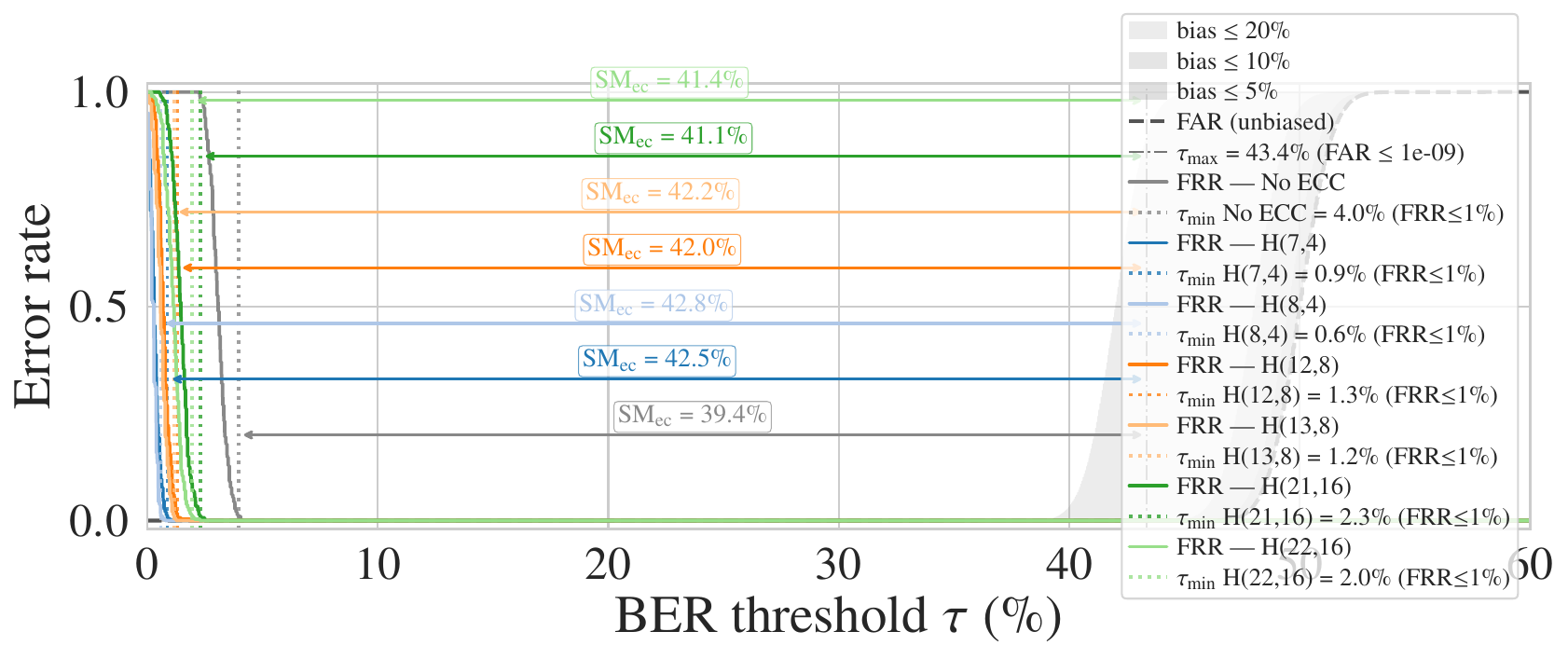}
        \caption{\ac{FAR} and \ac{FRR} vs. acceptance threshold analysis for $n=2048$.}
        \label{subfig:frr_1}
    \end{subfigure}

    \vspace{0.5cm}

    \begin{subfigure}[b]{\linewidth}
        \centering
        \includegraphics[width=\linewidth]{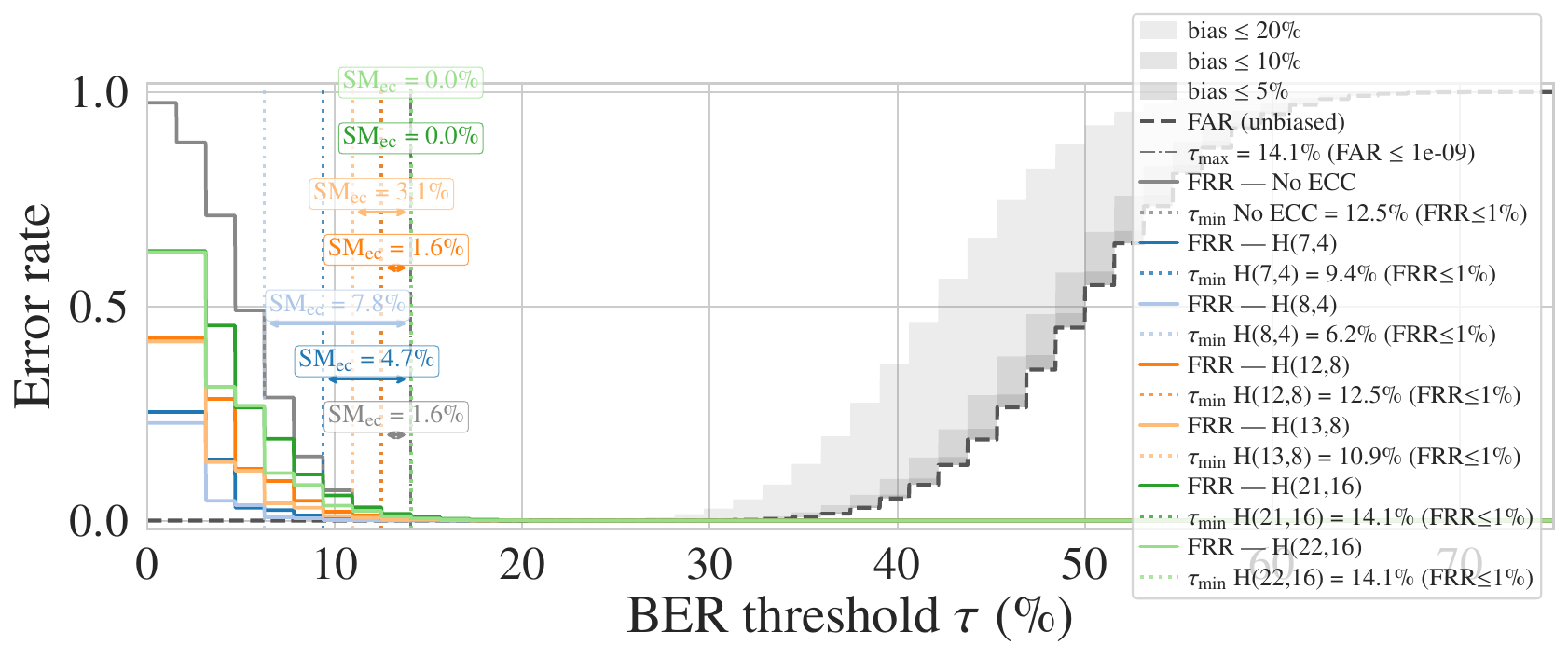}
        \caption{\ac{FAR} and \ac{FRR} vs. acceptance threshold analysis for $n=64$.}
        \label{subfig:frr_10}
    \end{subfigure}

    \caption{\ac{FAR} and \ac{FRR} vs. acceptance threshold analysis without \ac{mv}. 
    The graphs (\subref{subfig:frr_1}) ($n=2048$) and (\subref{subfig:frr_10}) ($n=64$) illustrate the trade-off between security (preventing unauthorized access, \ac{FAR}) and reliability (ensuring legitimate access, \ac{FRR}) as the acceptance threshold \(\tau_{\text{BER}}\) is varied. $\mathrm{SM}_\mathrm{ec}$ analysis is based on $\alpha_\mathrm{FAR} = 10^{-9}$ and $\alpha_\mathrm{FRR} = 0.01$.
    }
    \label{fig:FRR_plot}
\end{figure}

\noindent This section is about meeting real-world constraints via the selection of a device-specific authentication threshold \(\tau_{\text{BER}}\) that balances \ac{FAR} and \ac{FRR} while simultaneously quantifying the available design budget for resource optimizations for a fixed \ac{ec} and \ac{mv} setting.
The \ac{FAR} and \ac{FRR} represent the core security-usability trade-off in threshold-based authentication systems~\cite{1262027}. 
In \ac{HD}-based threshold systems,
as the \ac{FAR} is lowered (increasing security) by decreasing the threshold, the \ac{FRR} increases (reducing user convenience), creating a direct inverse relationship~\cite{3705677.3705691}. 
Consequently, for every \ac{ec} configuration the \ac{FAR} and \ac{FRR} are assessed in Fig.~\ref{fig:FRR_plot} based on the real-world genuine authentication measurement results (\ac{FRR}) and an analytical impostor \ac{puf} response (\ac{FAR}) distribution for varying $\tau_{\text{BER}}$ acceptance thresholds.

The plots in Fig.~\ref{fig:FRR_plot} highlight these two different types of errors, which move in opposite directions.
The y-axis represents the probability of a legitimate user being rejected (\ac{FRR}) or an impostor being accepted (\ac{FAR}).
Fig.~\ref{fig:FRR_plot} presents the results based on sweeping through \(\tau_{\text{BER}}\) from an extremely strict to an extremely lenient authentication acceptance threshold. 
In this analysis, the impostor \ac{puf} response and the associated \ac{BER} distribution after authentication are modeled analytically for a $n$-bit \ac{sram} \ac{puf} response.
Assuming ideal cell-to-cell independence and a uniform bit probability $p = 0.5$, the number of bit-flips between two independent responses follows a Binomial distribution $X \sim B(n, p)$. 
As the bit length $n$ of the \ac{puf} response increases, the spread ($\sigma$) of the binomial distribution for the impostor narrows significantly relative to the mean and vice versa.
Fig.~\ref{fig:FRR_plot} visually highlights this, because if a high enough \ac{puf} response bit-length (e.g.,~$n=2048$ in Fig.~\ref{subfig:frr_1}) is selected, the analytical impostor distribution exhibits extreme concentration around the mean.  
Consequently, this shows as a wide margin of separation between the genuine empirical \ac{FRR} and the theoretical \ac{FAR} curves in Fig.~\ref{fig:FRR_plot}.  
To visually demonstrate how bias effects shift the impostor distribution, 
models with slight bit biases $p \neq 0.5$ are also included in the analysis in Fig.~\ref{fig:FRR_plot}.
Once the threshold grows beyond the mean 
almost every theoretical impostor \ac{puf} response from the binomial model is considered authentic because random chance satisfies that requirement.
This analytical model is essentially the same concept as in~\cite{3705677.3705691}. 
In contrast, they introduce their analytical models using pre-measured mean reliability and uniqueness of their \ac{sram} \ac{puf}.
\ac{sram} \acp{puf} sufficiently approximate this binomial model~\cite{3705677.3705691}, as also highlighted in the previous uniformity results (see Fig.~\ref{fig:unif_hist}) and in the other close-to-ideal \ac{sram} \ac{puf} qualities observed in related work~\cite{4674345, 7127360}.
In Fig.~\ref{fig:FRR_plot}, this work employs this analytical binomial model only for the \ac{FAR} analysis (the ideal impostor) but uses actual empirical measurement curves for the \ac{FRR} (the legitimate user).
The analytical model just serves as a reference for ideal randomness and uniformity, against which the measured genuine noise distributions of the authentication iterations are compared.
We consider this approach as sufficient, as the focus of this work is not empirical inter-device \ac{FAR} security testing, but the resource-reliability trade-off.
Here, the binomial impostor model provides a clean, reproducible \ac{FAR} security baseline that isolates the variable this work cares about: how \ac{ec} and \ac{mv} configurations shift the \ac{FRR} curve relative to a fixed theoretical \ac{FAR} reference bound. 
Nevertheless, future analyses with different requirements may employ more complex mathematical models. 
These models could abandon ideal intra- and inter-device independence assumptions to account for real-world hardware realities, such as stuck bits, non-uniform cell probabilities, and spatial correlations between chips.

Visually, in Fig.~\ref{fig:FRR_plot}, the ideal operating window where the threshold should be placed is the flat gap between the \ac{FRR} lines hitting zero and the \ac{FAR} line starting to rise.
Within this gap, both the \ac{FRR} and \ac{FAR} are effectively zero, so the probabilities of an authentic node getting locked out or a malicious impostor node getting authenticated illegally are below a negligible threshold. However, these security guarantees warrant a formal definition.

This work aims to define the operating window for threshold selection using strict \ac{FAR} and \ac{FRR} error constraints based on the observed empirical \ac{FRR} and analytical \ac{FAR} values.
Let $\alpha_\text{FRR}$ be the maximum acceptable \ac{FRR}, and let $\alpha_\text{FAR}$ be the maximum acceptable \ac{FAR}.
This work defines a lower bound threshold, $\tau_{min}$, that satisfies the reliability constraint:
$$\operatorname{FRR}(\tau_\text{min}) \leq \alpha_\text{FRR}$$
Similarly, an upper bound threshold $\tau_{max}$ satisfies the security constraint against an impostor being accepted:
$$\operatorname{FAR}(\tau_\text{max}) \leq \alpha_\text{FAR}$$
Under the assumption of ideal bit independence and uniformity of the modeled impostor response, the probability of a false acceptance for a threshold placed anywhere in this operating range is strictly bounded by the designed-for error limit $\alpha_\text{FAR}$. 
By establishing strict upper limits for acceptable error rates $\alpha_\text{FAR}$ and $\alpha_\text{FRR}$, the boundary thresholds $\tau_{min}$ and $\tau_{max}$ that satisfy these constraints can then be numerically determined. 
In this analysis, $\alpha_\text{FRR}$ is conservatively set to $\alpha_\text{FRR} = 0.01$, as the limited sample count of the genuine \ac{BER} measurements is insufficient to reliably observe the rare outliers needed to define a stricter threshold.

An error-constrained security margin ($\mathrm{SM}_\text{ec}$) to quantify this operating window is then formally defined as the difference between these two bounded thresholds, representing the robust tolerance zone available for threshold calibration:
\begin{equation}
    \mathrm{SM}_\text{ec} = \tau_\text{max} - \tau_\text{min}
    \label{eq:sm_ec}
\end{equation}

If $\tau_{\text{min}} > \tau_{\text{max}}$, it indicates that the genuine \ac{puf} noise exceeds the maximum threshold permissible to block impostors. In this scenario, the system is fundamentally broken, as no threshold can satisfy both $\alpha_{\text{FRR}}$ and $\alpha_{\text{FAR}}$ simultaneously.
Therefore, a valid operating window strictly requires $\mathrm{SM}_{\text{ec}} > 0$. 
If $\mathrm{SM}_{\text{ec}} \le 0$, the magnitude of the negative margin quantifies the severity of the distribution overlap and serves as a diagnostic metric for failed \ac{puf}-assisted authentication primitives. 
When a valid margin exists, this work recommends selecting the acceptance threshold at the reliability boundary: $\tau_{\text{BER}} = \tau_{\text{min}}$. 
A potential violation of $\alpha_{\text{FRR}}$ is less severe, as it degrades user convenience but does not compromise the system's security guarantees.
If enough \acp{crp} are available, the legitimately rejected entity can reinitiate the authentication process with a fresh challenge from the verifier.
In addition, setting the threshold as strictly as possible maximizes the distance to $\tau_{\text{max}}$, thereby converting the entire $\mathrm{SM}_{\text{ec}}$ into a margin of safety against false acceptances.

Crucially, this work reframes $\mathrm{SM}_{\text{ec}}$ not just as a safety buffer, but as a quantifiable design budget.
Fig.~\ref{fig:FRR_plot} highlights $\mathrm{SM}_\text{ec}$ for   $\alpha_\text{FAR} = 10^{-9}$ and  $\alpha_\text{FRR} = 0.01$.
Depending on the ideal qualities of the \ac{puf} and these error thresholds,
a highly robust configuration (e.g., $n=2048$ with strong \ac{hc} \ac{ec}, see Fig.\ref{subfig:frr_1}) can yield a massively wide $\mathrm{SM}_{\text{ec}}$, indicating that the system is over-provisioned for the required $\alpha$ constraints. 
This excess margin reveals the potential headroom available for resource optimization. 
By actively optimizing the security margin, trading away excess $\mathrm{SM}_{\text{ec}}$ by reducing the \ac{puf} response length $n$ or utilizing less computationally demanding \ac{ec} and \ac{mv} schemes, designers can potentially significantly lower computational and energy overheads, or consequently opt for weaker hardware, while still strictly satisfying the designed-for security thresholds. 
If $\mathrm{SM}_\text{ec} = 0$, there is exactly one theoretical threshold that mathematically satisfies the constraints, but under real-world conditions a margin of zero means that environmental fluctuations, such as temperature changes, could break the security guarantees.
Therefore, a practical operating window only exists when $\mathrm{SM}_\text{ec} > 0$ whilst embedding some additional margin of safety.

Consequently, the engineering goal is to scale down the resource utilization
until a target~$\mathrm{SM}_{\text{ec}}^{\text{target}}$~is achieved, leaving only a small, deliberate safety buffer above zero to account for real-world environmental fluctuations that might otherwise break the security guarantees. A future algorithm to optimize the resource efficiency of the authentication scheme would solve for~$\mathrm{SM}_{\text{ec}}^{\text{target}}$~by locking acceptable error rates in place. 
This approach would mathematically optimize \ac{ec}, \ac{mv}, and \ac{puf} response length using the empirically determined correlations, actively altering the algorithmic parameters until the system perfectly fits the security requirements with minimal resource overhead (i.e.,~$\mathrm{SM}_{\text{ec}} \to \mathrm{SM}_{\text{ec}}^{\text{target}} \approx 0$).

\begin{figure}[!tb]
    \centering
    
    \begin{subfigure}[b]{0.49\linewidth}
        \centering
        \includegraphics[width=\linewidth]{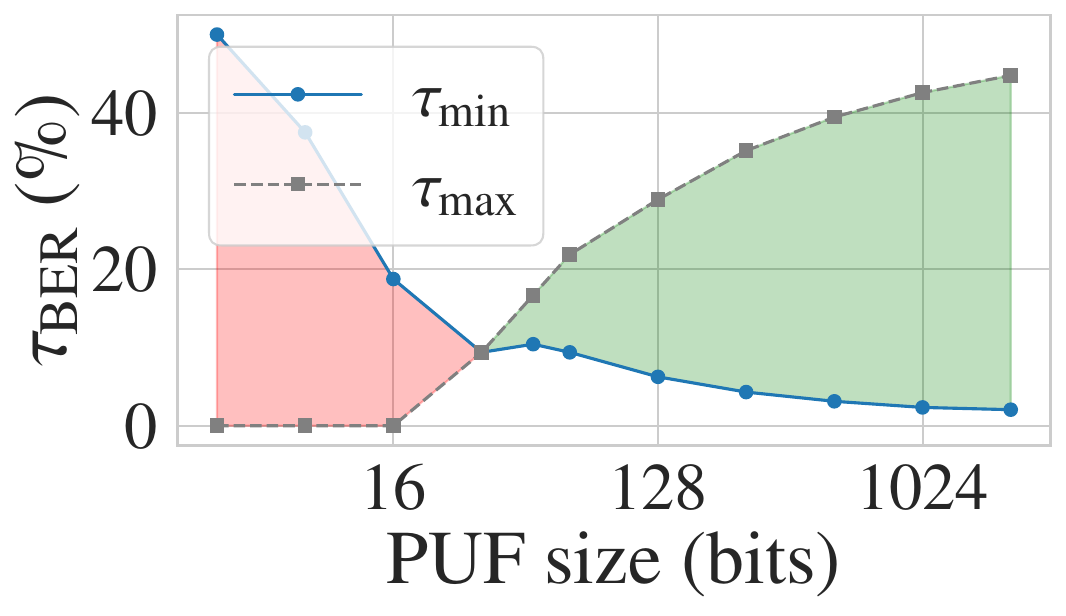}
        \caption{H(7,4), $N=1$, $\alpha_\text{FAR}=10^{-6}$.}
        \label{fig:subfig1}
    \end{subfigure}
    \hfill 
    \begin{subfigure}[b]{0.49\linewidth}
        \centering
        \includegraphics[width=\linewidth]{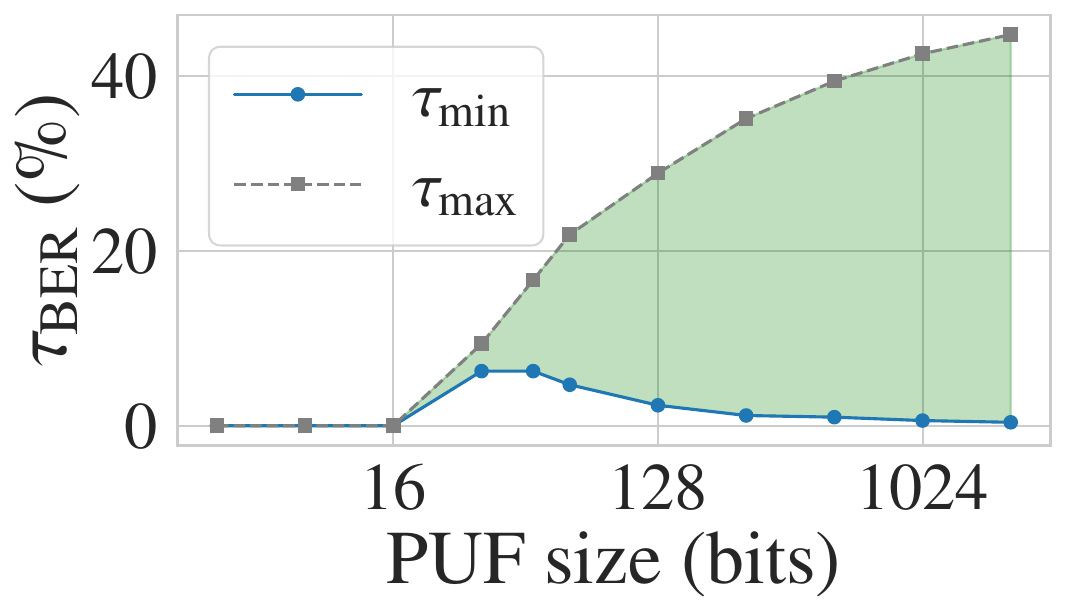}
        \caption{H(7,4), $N=20$, $\alpha_\text{FAR}=10^{-6}$.}
        \label{fig:subfig2}
    \end{subfigure}
    \vspace{1em}
    
    \begin{subfigure}{\columnwidth}
        \centering
        \includegraphics[width=\linewidth]{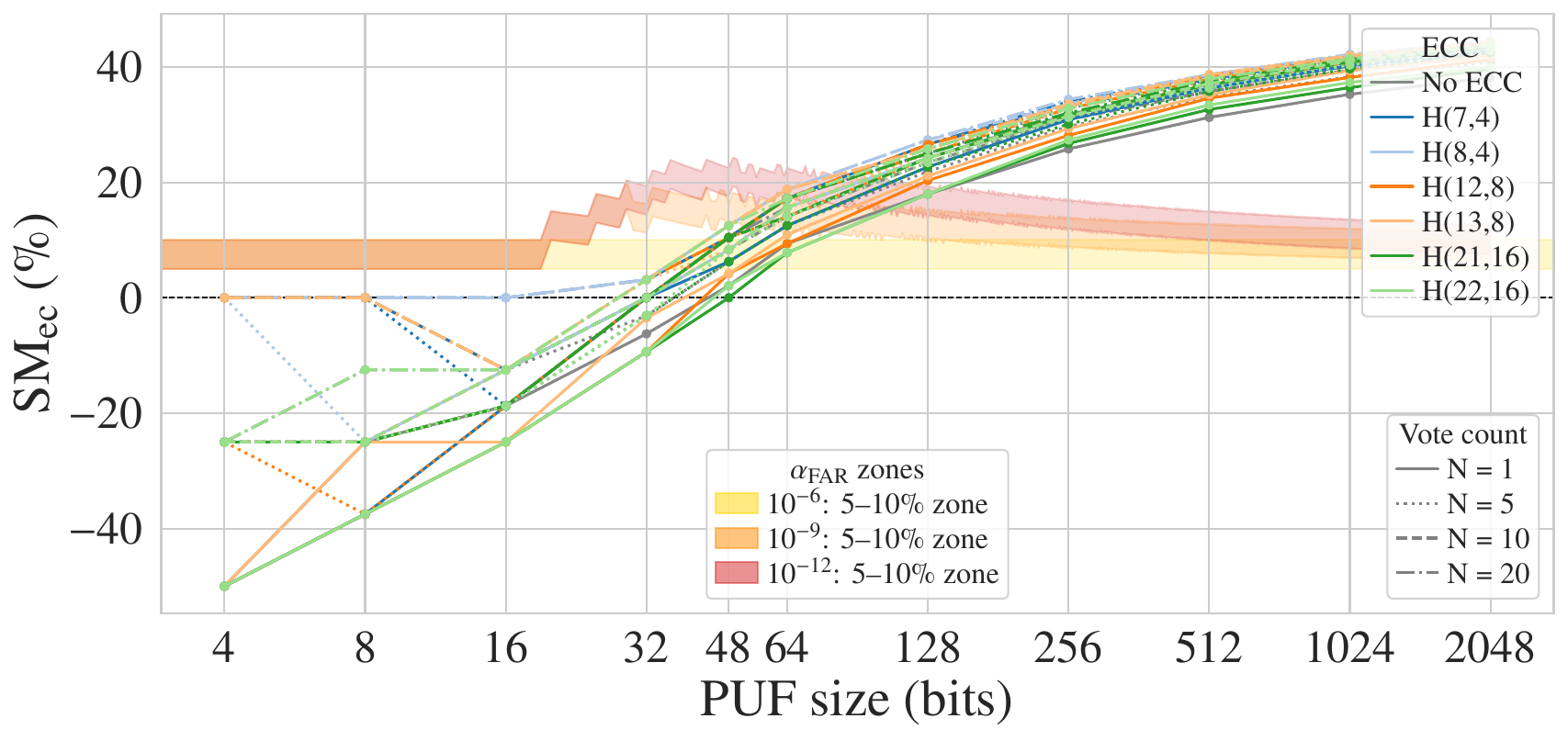}
        \caption{$\mathrm{SM}_{\text{ec}}$ scaling across all configurations vs. \ac{puf} size $n$.}
        \label{fig:subfig3}
    \end{subfigure}
    \vspace{1em}
    
    \begin{subfigure}{\columnwidth}
        \centering
        \includegraphics[width=\linewidth]{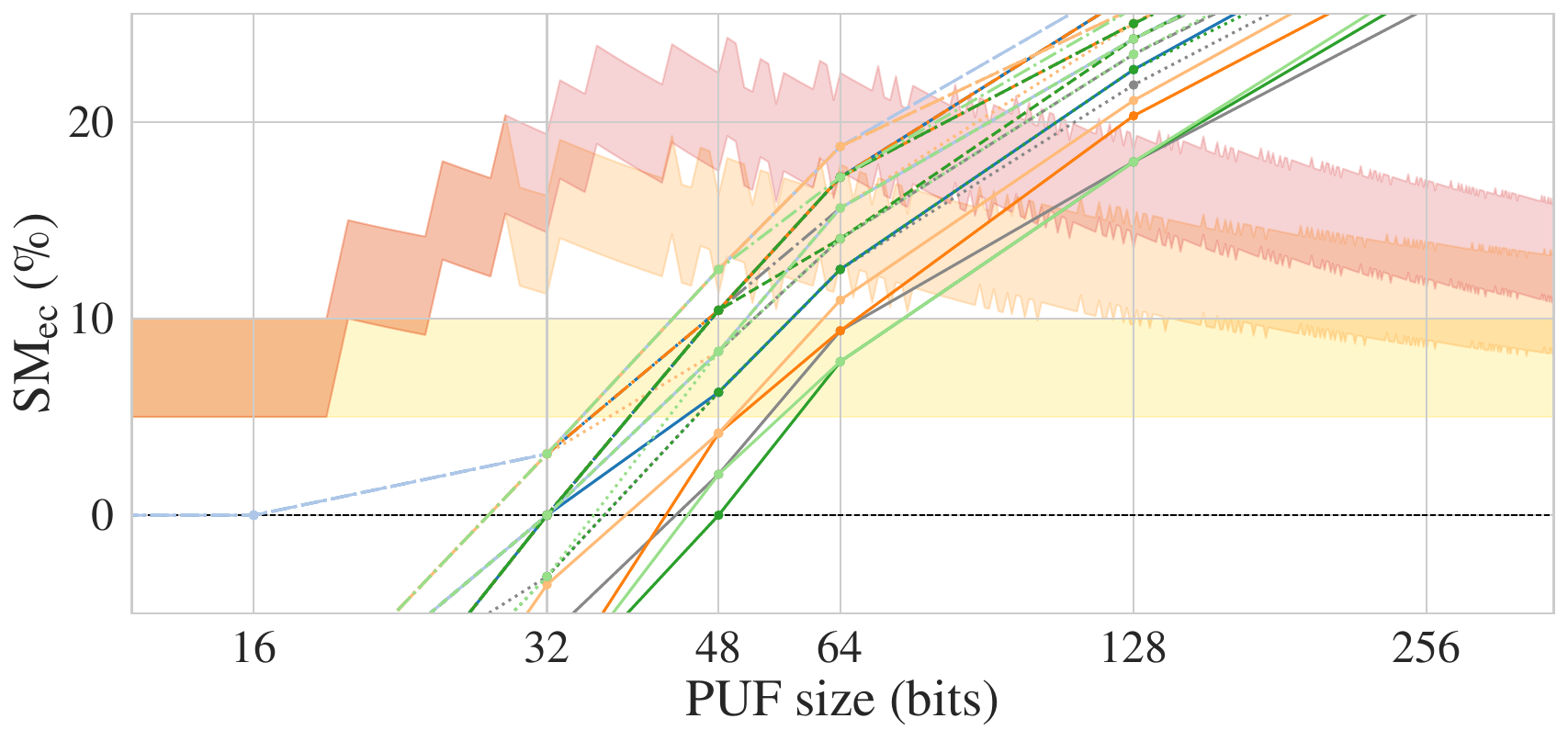}
        \caption{$\mathrm{SM}_{\text{ec}}$ scaling zoomed in across all configurations vs. \ac{puf} size $n$.}
        \label{fig:subfig4}
    \end{subfigure}
    
    \caption{
    Evolution of the error-constrained security margin ($\mathrm{SM}_{\text{ec}}$) at $\alpha_\mathrm{FAR} = 10^{-6}$. (\subref{fig:subfig1}) and (\subref{fig:subfig2}) show $\mathrm{SM}_{\text{ec}}$ bounded by $\tau_{\text{min}}$ and $\tau_{\text{max}}$ for an exemplary configuration. Green denotes valid operating windows ($\mathrm{SM}_{\text{ec}} > 0$) and red marks unviable configurations. 
    (\subref{fig:subfig3}) maps $\mathrm{SM}_{\text{ec}}$ scaling versus $n$, isolating configurations above a safe limit $\mathrm{SM}_{\text{ec}}^{\min} = 5\,\%$ where excess resource overhead is constrained by an acceptable upper bound in 5--10\% target $\alpha_\text{FAR}$ zones. 
    (\subref{fig:subfig4}) provides a closer view.
    }
    \label{fig:SM_ec_scaling}
\end{figure}

For now, the evaluations of the subsequent Sections offer a comparative summary of $\mathrm{SM}_\text{ec}$ for various configurations of the implemented authentication system for specific security targets $\alpha_{\text{FAR}}$. 
To start, Fig.~\ref{fig:SM_ec_scaling} illustrates how $\mathrm{SM}_{\text{ec}}$ scales with \ac{puf} size $n$ across all evaluated configurations at $\alpha_\mathrm{FAR} = 10^{-6}$.

\subsubsection{Relative Impact of MV, $n$, and HC EC}
All \ac{ec} and \ac{mv} configurations exhibit an increasingly robust security margin $\mathrm{SM}_\mathrm{ec}$ with increasing $n$, $N$ or decreasing \ac{hc} code rate. 
The magnitude of each parameter's contribution differs substantially:
At $\alpha_\mathrm{FAR} = 10^{-6}$ scaling $n$ from 64 to 2048\,bits improves $\mathrm{SM}_\text{ec}$ by
approximately 28\,\%, whereas maximising $N$ (from 1 to 20 votes) or
switching to the most redundant \ac{ecc} variant each add only $\approx\!6$\,\%
at $n\!=\!256$\,bits. 
This asymmetry arises because just by increasing $n$ the impostor \ac{BER} distribution becomes more concentrated, and the genuine empirical \ac{BER} distribution becomes tighter, shifting both $\tau_\text{max}$ and $\tau_\text{min}$ so that they separate more from each other and the $\mathrm{SM}_\text{ec}$ improves, whereas $N$ and \ac{hc} \ac{ec} affect $\tau_\text{min}$ alone.
Consequently, for a high $n$, the results show, that the law of large numbers works in an engineer's favor to create a nearly impassable statistical barrier for authentication, even for relatively lenient acceptance thresholds.
Nevertheless, Fig.~\ref{fig:SM_ec_scaling} highlights that increasing $n$ similarly to increasing $N$ or decreasing the \ac{hc} code rate at some point yields diminishing returns in increasing $\mathrm{SM}_{\text{ec}}$. 
Consequently, the combined use of all three might be necessary in practice for very strict error thresholds or when a small $n$ is dictated by resource restrictions.

\subsubsection{Impact of Tightened Security Constraints}
\label{subsubsec:changing_far_constraint}
\begin{figure}[t]
  \centering
  \includegraphics[width=\columnwidth]{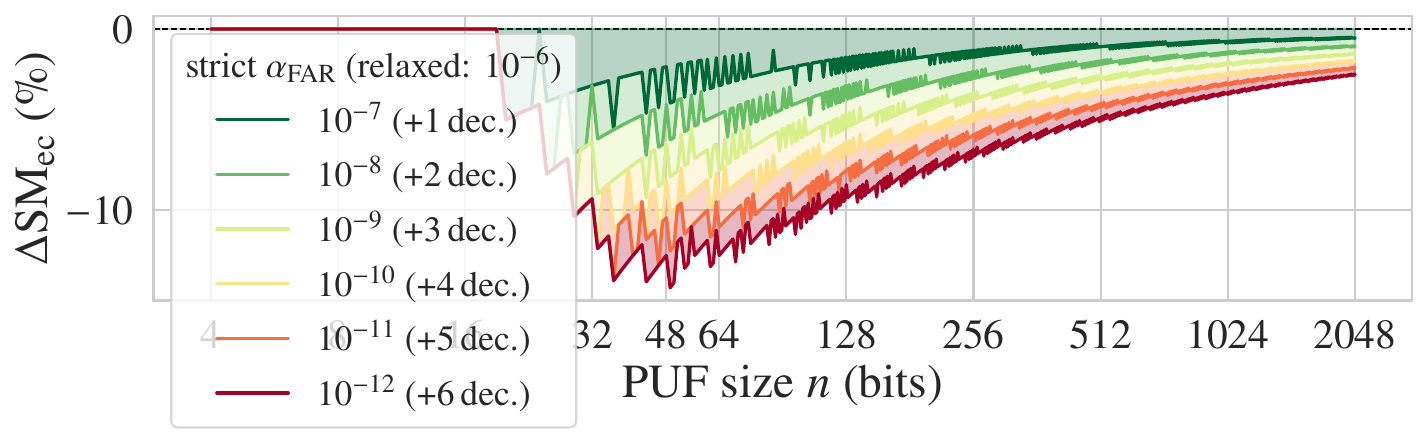}
  \caption{
    Change in $\Delta\text{SM}_\text{ec}$ when tightening $\alpha_\text{FAR}$ from $10^{-6}$, as a function of \ac{puf} size~$n$ for an ideal impostor with no bias.
  }
  \label{fig:delta-sm-ec}
\end{figure}

Fig.~\ref{fig:delta-sm-ec} plots the shift in the error-constrained security margin $\Delta\mathrm{SM}_\mathrm{ec}$ across \ac{puf} size $n$ as the $\alpha_{\text{FAR}}$ requirement is tightened.
Trivially, stricter security guarantees also reduce $\mathrm{SM}_\text{ec}$ leaving less potential for resource optimizations.
Fig.~\ref{fig:delta-sm-ec} highlights that making $\alpha_{\text{FAR}}$ stricter (e.g., $10^{-6} \rightarrow 10^{-9}$) has a greater effect on $\mathrm{SM}_\mathrm{ec}$ at smaller values of $n$. 
For example, for $2048$-bit measurements, the choice between $\alpha_{\text{FAR}} = 10^{-6}$ and $\alpha_{\text{FAR}} = 10^{-9}$ makes little difference ($\approx 1.4\%$).
However, for $64$ to $128$\,bit responses, the difference is significant.
The oscillations are a quantisation artefact: as $n$ grows, the integer number of bits separating the two acceptance thresholds can only change in unit steps, so $\Delta\mathrm{SM}_\mathrm{ec}$ falls gradually between consecutive steps and rises sharply at each forward step, tracing a sawtooth whose amplitude envelope decays as $1/\sqrt{n}$. 
At $n \leq 16$, the Binomial impostor distribution has so few discrete outcomes that $\tau_\text{max}$ floors at 0\,\% under all $\alpha_\text{FAR}$ targets, making $\Delta\mathrm{SM}_\mathrm{ec}$ identically zero. 
The \ac{puf} response is too short for the \ac{FAR} constraint to differentiate between the security levels. 
When $\alpha_{\text{FAR}}$ is made stricter $\tau_\mathrm{max}$ shifts, and the $\mathrm{SM}_\mathrm{ec}$ is reduced by a fixed amount at a given $n$. 
Because an ideal binomial impostor source is used $\Delta\mathrm{SM}_\mathrm{ec}$ can also be analytically determined.
When $\alpha_{\text{FRR}}$ is tightened, the resulting $\Delta\mathrm{SM}_{\text{ec}}$ values are inherently less predictable than in the $\alpha_{\text{FAR}}$ case,
as in this work, $\tau_{\min}$ is derived from empirical genuine \ac{BER} measurements with finite sample counts rather than a closed-form model. 
Nevertheless, assuming the genuine \ac{BER} also follows a binomial distribution as shown in~\cite{3705677.3705691}, $\Delta\mathrm{SM}_{\text{ec}}$ behaves analogously for $\Delta\alpha_{\text{FRR}}$ and $\Delta\alpha_{\text{FAR}}$, 
so the analysis presented for the \ac{FAR} extends qualitatively to the \ac{FRR}.

\subsubsection{PUF Bias Effects on Threshold Calibration}
\label{subsubsec:puf_bias_effect}
\begin{figure}[t]
  \centering
    \includegraphics[width=\columnwidth]{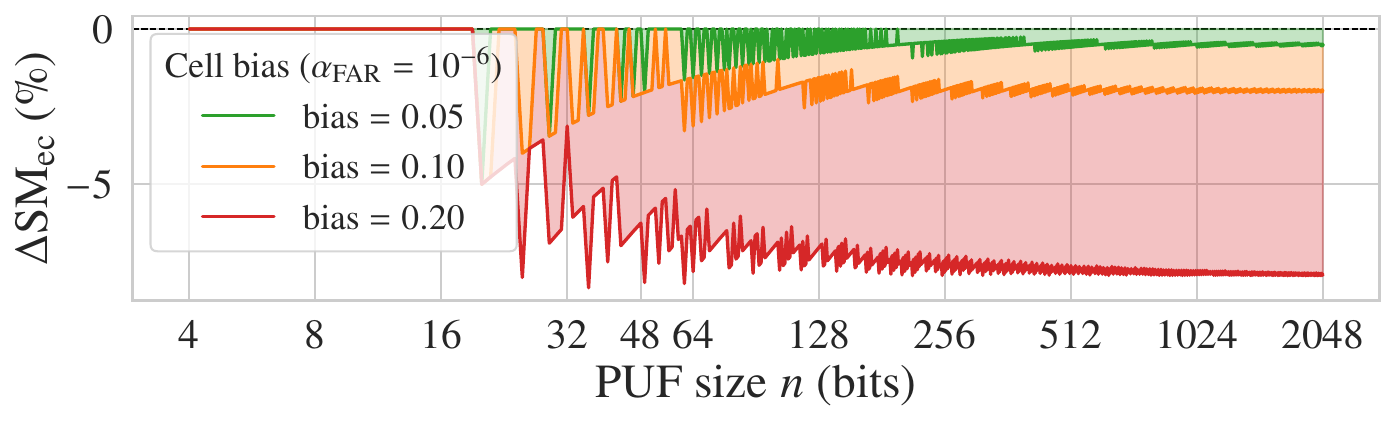}
  \caption{
Change in $\Delta\text{SM}_\text{ec}$ for non-ideal bit bias.
  }
  \label{fig:delta-bias}
\end{figure}

Fig.~\ref{fig:delta-bias} shows that $\mathrm{SM}_\text{ec}$ is highly dependent on the ideal qualities of the \ac{puf} response, as $\mathrm{SM}_\text{ec}$ degrades with increased bit bias in the binomial impostor model. 
The sawtooth pattern arises due to similar reasons as in~\ref{subsubsec:changing_far_constraint}.

\subsubsection{Imperfect PUF Uniqueness}
\label{subsubsec:puf_corr_effect}
\begin{figure}[t]
  \centering
    \includegraphics[width=\columnwidth]{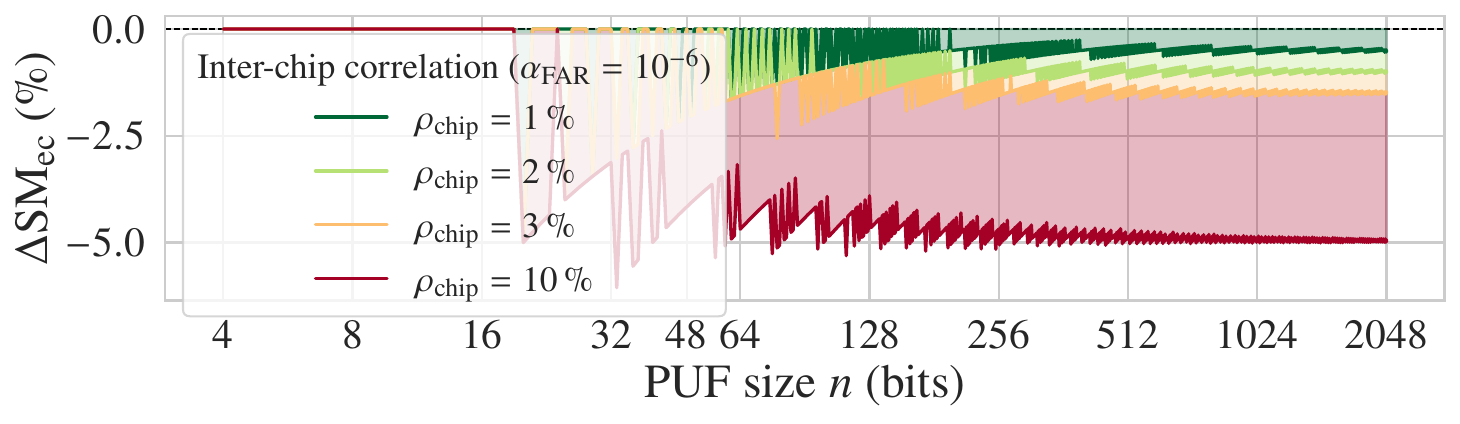}
  \caption{
  Impact of inter-chip correlation $\rho_{\mathrm{chip}}$ on $\mathrm{SM}_\text{ec}$. 
  }
  \label{fig:interchip-corr-sweep}
\end{figure}

Similarly to~\ref{subsubsec:puf_bias_effect}, imperfect \ac{puf} uniqueness caused by inter-chip correlation reduces $\mathrm{SM}_{\mathrm{ec}}$. 
The impostor inter-chip \ac{HD} is binomially modeled, where the baseline per-bit mismatch probability is proportionally reduced by a positive inter-chip correlation factor $\rho_{\mathrm{chip}} \in [0,1)$. 
Because correlated chips generate inherently more similar responses, the theoretical impostor \ac{HD} distribution shifts closer to the genuine authentication distribution. Consequently, to mathematically maintain a strict target ($\alpha_\text{FAR}$), the maximum permissible impostor threshold $\tau_{\max}$ must be lowered. 
Fig.~\ref{fig:interchip-corr-sweep} illustrates this correlation-induced shift $\Delta\mathrm{SM}_\text{ec}$, plotting the difference between an uncorrelated and a correlated baseline.
As expected, it shows, a larger correlation factor $\rho_{\mathrm{chip}}$ necessitates a stricter threshold, yielding a larger loss in $\mathrm{SM}_\text{ec}$.
The characteristic sawtooth behavior again arises from the discrete nature of the binomial quantile at finite \ac{puf} sizes $n$.

\subsubsection{Excluding Overprovisioned Configurations}
\label{subsubsec:excluding_configs}
Out of all tested configurations in Fig.~\ref{fig:SM_ec_scaling}, for a given reliability constraint ($\alpha_\text{FRR} = 0.01$) and security constraint (e.g., $\alpha_\text{FAR} = 10^{-6}$), the configurations within the shaded target bands are deemed most appropriate, as resource overhead is strictly capped to prevent overprovisioning.
These bands isolate an operational window where the security margin is bounded strictly between a necessary safety floor ($\mathrm{SM}_{\text{ec}}^{\min} \geq 0$) against unaccounted for \ac{puf} noise and inter-chip dependencies, and an acceptable overhead ceiling $\mathrm{SM}_{\text{ec}}^{\text{ceil}}$. 
Configurations within this zone safely satisfy all error constraints while exhibiting little wasteful headroom for further resource optimization.
Although visualized here as a broader selection band ($\mathrm{SM}_{\text{ec}}^{\min} \leq \mathrm{SM}_{\text{ec}}^{\text{target}} \leq \mathrm{SM}_{\text{ec}}^{\text{ceil}}$) to illustrate the available design space, a practical deployment would aggressively optimize resources by collapsing this target directly to the safety floor ($\mathrm{SM}_{\text{ec}}^{\text{target}} = \mathrm{SM}_{\text{ec}}^{\min}$).

To account for non-ideal \ac{puf} behavior that was not empirically assessed in this work, for instance, correlations between chips (see~\ref{subsubsec:puf_corr_effect}), a fixed safety offset of $\mathrm{SM}_{\text{ec}}^{\min} = 5\,\%$ is applied. 
This is assumed to be sufficient, due to the close-to-ideal uniformity results of the implemented \ac{sram} \ac{puf} (see Fig.~\ref{fig:unif_hist}) and predictably good \ac{sram} \ac{puf} uniqueness and randomness as shown in related work~\cite{3705677.3705691, 4674345, 7127360}.
But this safety offset has to be individually assessed based on the \ac{puf}.

Statistically, $\tau_{\max}$ remains zero at small $n$ until $\alpha_{\text{FAR}} > 0.5^n$ (approximately $n \gtrsim 20$ for $\alpha_{\text{FAR}} = 10^{-6}$ and $n \gtrsim 30$ for $\alpha_{\text{FAR}} = 10^{-9}$), marking the theoretical threshold beyond which a strictly positive security margin ($\mathrm{SM}_{\text{ec}}>0$) becomes possible.
At small $n$ close to this bound, a counter-intuitive non-monotonic $\mathrm{SM}_{\text{ec}}$ inversion can be observed in Fig.~\ref{fig:SM_ec_scaling} for some of the \ac{hc} \ac{ec} and \ac{mv} configurations. 
This effect is due to evaluating a statistical quantile ($\tau_\text{min}$ derived from $\alpha_\text{FRR}$) against an asymmetric, multi-modal distribution of genuine post-authentication \ac{BER}, caused by rare clustered errors that increasingly start to fragment the distribution at smaller $n$.
Decreasing $n$ increasingly isolates clustered erroneous bits into a shrinking fraction of rare worst-case outlier responses, leaving a proportionally larger pool of perfect, error-free ones.
This creates competing effects: while the increased dispersion of the \ac{BER} distribution at lower $n$ (see Fig.~\ref{fig:ber_vs_votes}) initially drives $\tau_{\min}$ up, the growing sample pool of perfect responses causes the percentile thresholding at the $\alpha_\text{FRR}$ 
boundary to increasingly mask the rare outliers in some cases. 
In these cases, once this masking effect overpowers the increasing dispersion, the $\tau_{\min}$ threshold counter-intuitively stops increasing or is even pulled back towards zero. 
This statistical effect can be amplified by \ac{hc} miscorrections because these induce additional errors in a codeword, making the effect less noticeable in the no-\ac{ecc} case.
If errors were truly randomly scattered across bits, smaller $n$ would always yield a wider non-fragmented \ac{BER} distribution.
The 99th percentile ($\tau_{min}$ for $\alpha_\text{FRR} = 0.01$) would then always decrease monotonically as $n$ grows.

Consequently, once the growing fragmentation of the per-response \ac{BER} distribution starts to have a stronger impact on the $\tau_\text{min}$ computation with decreasing $n$, this effect increasingly degrades the validity of $\mathrm{SM}_{\text{ec}}$. 
Therefore, $\mathrm{SM}_{\text{ec}}$ is only a reliable indicator of a resource-aware and error-constrained operating window available for threshold calibration above a lower bound $n_\text{min}$.
Based on a visual inspection of Fig.~\ref{fig:SM_ec_scaling}, rather than a formal empirical assessment, this bound $n_\text{min}$ appears to lie at approximately $n \gtrsim 64$ for $\alpha_\text{FAR} = 10^{-6}$, beyond which this effect ceases to have a noticeable impact.
Since $n_\text{min}$ strongly depends on the magnitude of \ac{puf} error clustering and the nature of the \ac{BER} distribution at small $n$ it is in need to be individually assessed.
Therefore, only configurations in Fig.~\ref{fig:SM_ec_scaling} that exceed $n \gtrsim n_\text{min}$ and fall within a specific $\alpha_\text{FAR}$ target zone are considered candidates for a resource-aware $\tau_\text{BER}$ calibration. 
Below this threshold, this work advises against relying on $\mathrm{SM}_{\text{ec}}$, as the metric becomes severely distorted by the exact shape of the distribution, 
a characteristic assumed to be highly device-specific and therefore unreliable when used to assess whether a \ac{sram} \ac{puf}-assisted authentication schemes reliably meets certain $\alpha_\text{FAR}$ and $\alpha_\text{FRR}$ targets.

For a shift in the security constraint $\alpha_\mathrm{FAR}$ the offsets $\Delta\mathrm{SM}_{\text{ec}}$ in Fig.~\ref{fig:delta-sm-ec} would need to be applied to the $\mathrm{SM}_{\text{ec}}$ curves in Fig.~\ref{fig:SM_ec_scaling}.
Instead, each additional $\alpha_\mathrm{FAR}$ target zone in Fig.~\ref{fig:SM_ec_scaling} is reconstructed analytically by shifting the baseline band by the $\tau_{\max}$ difference between baseline and target constraints, without re-plotting the empirical $\text{SM}_{\text{ec}}$  lines.
For configurations within these target zones, the trade-offs among $n$, $N$, and the \ac{hc} \ac{ec} variant, such as the computational overhead discussed in Section~\ref{subsec:Design_Space_Exploration}, now dictate the final selection.

\subsubsection{Implications of the Results}
The target zones in Fig.~\ref{fig:SM_ec_scaling} indicate,
designers can opt for a larger \ac{puf} response $n$ and relax the acceptance threshold instead of resorting to implementing complex and computationally demanding \ac{hc} \ac{ec} or \ac{mv} schemes, whilst meeting the same security guarantees at a designed-for $\mathrm{SM}_{\text{ec}}^{\text{target}}$. 
Only shorter \ac{puf} responses thus necessitate stronger \ac{mv} and \ac{hc} \ac{ec}. 
Crucially, as Fig.~\ref{fig:SM_ec_scaling} shows, \ac{hc} \ac{ec} can even worsen the security margin compared to the baseline at smaller $n < 128$. 
Only after $n > 128$ this apparent performance inversion is reliably reverted and Fig.\ref{fig:SM_ec_scaling} shows the expected pattern of \acp{hc} with lower codes rates always outperforming higher code rates.
This is in contrast to the average \ac{BER} which reliably improves with the code rate independent of $n$ (see Fig.\ref{fig:ber_vs_votes}).
With fewer \ac{hc} codewords per \ac{puf} response, due to smaller $n$, a single miscorrected block or other clustered errors disproportionately affect the per-response \ac{BER}, creating worse outliers than without \ac{ec} (see also Fig.~\ref{fig:ber_vs_votes}), which can lead to a significant fragmentation of the \ac{BER} distribution.
For larger \ac{puf} sizes, the correctly decoded blocks, where single-bit errors are eliminated, increasingly outweigh the rare miscorrections, and the net effect of \ac{ec} becomes reliably positive.
Again due to the percentile thresholding computation explained in Section~\ref{subsubsec:excluding_configs}, this results in no-\ac{ecc} actually reliably outperforming some \ac{hc} variants in meeting $\mathrm{SM}_{\text{ec}}^{\text{target}}$ at smaller $n$ (for example, see H(21,16) in Fig~\ref{fig:SM_ec_scaling} at $n=64$).
Consequently, at small values of $n$, \ac{hc} configurations with efficient code rates (e.g., H(21,16) or H(22,16)) are often unviable. This makes it difficult to justify \ac{hc} deployment over increasing $n$, which naturally improves the security margin and renders such \ac{puf} stabilization techniques increasingly obsolete.
For \ac{mv}, the main drawback remains the computational overhead discussed in Section~\ref{subsec:Design_Space_Exploration}.
This makes small $n$ only attractive in edge cases,
when a large $n$ is prohibitively expensive, due to an increasingly large \ac{crp} database, \ac{nvs} limitations, wireless transmission overheads, or other constraints.
Here, the results highlight that the system can be made significantly stricter without degrading usability through one of three approaches: moderate \ac{mv} ($N\lesssim5$) combined with low-redundancy \acp{hc}, low code rate \ac{hc} \ac{ec} alone, or aggressive \ac{mv} ($N \gtrsim 5$).
Consequently, only for a small $n$, selecting the appropriate reliability optimization in threshold-based \ac{puf}-assisted authentication remains an application and device-specific challenge.
To this end, this implementation presented a representative assortment of resource-aware solutions. 
However, designers must also balance the quality of the selected reliability enhancement against the mentioned security implications of increased helper data storage~\cite{10.1145/3591464, 8254007} and the computational and memory overheads discussed in Section~\ref{subsec:Design_Space_Exploration}. 
Combined, these diverse factors introduce a multitude of competing dimensions in this design space exploration that engineers must consider when deciding on an authentication configuration.
Consequently, to satisfy the strict resource constraints of \ac{iiot} devices, this work recommends a balanced authentication approach.
Here, the authors assume the perfect trade-off to be highly device and application-unique and in need to be individually assessed. 
Specifically, whenever a large~$n$~is feasible, 
to meet security guarantees in a resource-efficient way~(i.e.,~$\mathrm{SM}_{\text{ec}} \to \mathrm{SM}_{\text{ec}}^{\text{target}} \approx 0$) it is preferred to opt for a more lenient acceptance threshold at such larger $n$, and when necessary, pair it with a lightweight \ac{hc} \ac{secded}, rather than high \ac{mv} counts.

\section{Conclusion and Outlook}
\label{sec:conclusion}
\noindent This work presented a highly secure threshold-based \ac{sram} \ac{puf}-assisted authentication scheme tailored for resource-constrained \ac{iiot} environments. 
To this end, it formally defined an error-constrained security margin ($\mathrm{SM}_\text{ec}$) to quantify the operating window available for threshold calibration between the strict boundaries of acceptable false acceptance (\ac{FAR}) and rejection rates (\ac{FRR}).
This novel metric was systematically utilized as a measurable design budget, highlighting the potential to safely scale down computational overhead without violating predefined security and reliability guarantees.
It depends heavily on ideal \ac{sram} \ac{puf} qualities, and larger responses ($n$) permit significantly more lenient acceptance thresholds, rendering reliability optimizations increasingly obsolete.

The evaluation demonstrated the viability of using \ac{mv} paired with low-overhead \acp{hc} to effectively 
maintain \ac{FRR} compliance under strict thresholds,
particularly when resource limits mandate shorter \ac{puf} responses.
All evaluated \ac{ec} variants exhibited a monotonic mean per-response post-authentication \ac{BER} reduction with increasing \ac{mv} count and decreasing \ac{hc} \ac{ec} code rate.
More redundant and computationally expensive combined configurations reliably cap the mean \ac{BER} below 1\%, with both strategies yielding diminishing returns, but increasingly prohibitive overheads.
\ac{secded} \acp{hc} exhibit only marginally better mean \ac{BER}, but the results show that increased \ac{sec} miscorrection-induced outliers, manifesting in an increased fragmentation of the genuine \ac{BER} distribution, can actually significantly worsen the $\mathrm{SM}_\text{ec}$ at smaller $n$ even when compared to the base case. 
This suggests \ac{secded} deployment is preferred at small $n$,  even though it must be carefully evaluated against the additional parity overhead on a per-application basis.

Combined, these results facilitate a robust design space exploration of various \ac{sram} \ac{puf}-assisted authentication configurations, 
guiding future implementations in trading off excess $\mathrm{SM}_\text{ec}$ against processing time, energy requirements, and implementation overhead.
Here, as a first step in this direction, this work clearly demonstrated \ac{ec} quality and computational overheads of \ac{mv} and \acp{hc}. 

Future work could expand upon this by incorporating detailed power and energy profiling, multi-board empirical \ac{FAR} testing and more complex mathematical impostor models.
 
\section*{Acknowledgment}
{
\hyphenpenalty=10000
\exhyphenpenalty=10000
\noindent This research was conducted within the ALPAKA and \mbox{SUSTAINET\_guarDian} research project, funded by the German Federal Ministry of Research, Technology and Space (BMFTR) under the grant 16KIS1841K and 16KIS2239K.
\par
} 
 
\small 
\bibliographystyle{unsrt}
\bibliography{references.bib}

\end{document}